\documentclass[12pt]{iopart}
\usepackage{graphicx}
\usepackage{iopams}  

\begin{document}

\title{Nonlinear entanglement witnesses for four qubits in mutually unbiased bases}

\author{K. Aghayar}
\address{Physics Department, Urmia University, 11 km Sero Road, Urmia, Iran }
\ead{k.aghayar@urmia.ac.ir}

\author{A. Heshmati}
\address{Department of Physics, Shabestar Branch, Islamic Azad University, Shabestar, Iran}
\ead{heshmati@tabrizu.ac.ir}

\author{M. A. Jafarizadeh}
\address{Department of Theoretical Physics and Astrophysics, University of Tabriz, Tabriz, Iran}
\ead{jafarizadeh@tabrizu.ac.ir}

\vspace{10pt}
\begin{indented}
\item[]February 2020
\end{indented}

\begin{abstract}
	Entanglement witness is a Hermitian operator that is useful for detecting the genuine multipartite entanglement of mixed states. Nonlinear entanglement witnesses have the advantage of a wider detection range in the entangled region. We construct genuine entanglement witnesses for four qubits density matrices in the mutually unbiased basis. First, we find the convex feasible region with positive partial transpose states. Then, to reveal the entangled regions, we present some appropriate linear entanglement witnesses and, we find the envelope of this family of linear witnesses as a nonlinear witness. Finally, we study thermal entanglement and we show for some Hamiltonians the witnesses can detect the entanglement at all temperatures.
\end{abstract}

%
\vspace{2pc}

\section{Introduction}\label{intro}

Quantum entanglement is a physical phenomenon that occurs when pairs or groups of particles are generated or interact in ways such that the quantum state of each particle cannot be described independently instead, a quantum state may be given for the system as a whole \cite{RefJ:Horodecki_2009,RefJ:Raimond}. Mathematically, a state of a composite quantum system is called entangled if it cannot be written as a convex combination of product states \cite{RefJ:Werner_1989}. Quantum entanglement has many physical applications such as quantum key distribution in quantum cryptography \cite{RefJ:Ekert_1991,RefJ:Ursin} with new experiments \cite{RefJ:Bunandar,RefJ:Sasaki}, quantum dense coding \cite{RefJ:Nepal}, and quantum teleportation \cite{RefJ:Bennett,RefJ:Wang}. In these applications, there must be some physical observable acting on system state to detect the entanglement in the system. One of the observable detecting entanglement especially for a system with three or more particles is entanglement witness ($ EW $).
\par
Entanglement witness is an observable which completely characterize separable states and detect entanglement in a system experimentally \cite{RefJ:Bourennane,RefJ:Monteiro,RefJ:Horodecki_1996,RefJ:Terhal_2000,RefJ:Koh}. From a geometrical point of view as the quantum mixed state family (density matrices) is a convex set so an $ EW $ can be described by hyperplanes in the density matrix space. Now the $ EW $, $ W $ is a Hermitian operator with non-negative value on all pure product states
$
Tr( W \rho_{_{Product}}) \geq 0
$
where $ \rho_{_{Product}} = | \psi_{1} \rangle ... | \psi_{n} \rangle \langle \psi_{n} | ... \langle \psi_{1} | $. The entanglement of $ \rho $ is detected by $ EW $ if and only if
$
Tr( W \rho ) \leq 0
$.
Although there is a necessary and sufficient condition for separability in $ 2 \otimes 2 $ and $ 2 \otimes 3 $ cases, called the positive partial transpose $ PPT $ criterion or Peres-Horodecki criterion \cite{RefJ:Horodecki_1996}, in general, there is no such condition for other cases and there are states that are entangled but $ PPT $ in all those cases which are called $ PPT $ entangled states. The other way for detecting entanglement for systems with higher dimensions is by using $ EW $. Especially the $ EW $s detecting $ PPT $ entangled states are of great importance. Usually these are non-decomposable $ EW $s or optimal $ EW $s. One can consider linear $ EW $s which is relatively simple to construct or nonlinear $ EW $s.
\par
There are some nonlinear separability criteria in the literature. Generally, these nonlinear $ EW $s have a wider range of entanglement detection. In the article \cite{RefJ:Guhne_2004}, the author derives a family of necessary separability criteria for finite-dimensional systems based on inequalities for variances of observables and formulate an equivalent criterion in terms of covariance matrices. The criteria may be applied from the regime of continuous variables to finite-dimensional systems. Nonlinear \emph{EW}s as an extension of linear witnesses with the ability to detect the states with negative partial transpose has been presented in \cite{RefJ:Guhne_2006}. A general theorem as a necessary condition for the separability based on concave-function uncertainty relations has been derived for both finite and infinite-dimensional systems in \cite{RefJ:Huang_2010}. The author has been using the specific concave function method for a system with mutually unbiased bases (MUB), for entanglement detection as a special case of his approach. In some cases that approach leads to an analytic entanglement detection which is stronger than the Shannon entropy uncertainty relation and the Landau-Pollak uncertainty relation. Using an appropriate class of uncertainty relations, the entanglement of the local quantum states of a pair of $ N $-level systems have been defined in \cite{RefJ:Hofmann}. These uncertainty relations may be used as an experimental test of entanglement generation. A derivation of nonlinear \emph{EW}s based on covariance matrices has been investigated in \cite{RefJ:Guhne_2007}. The nonlinear functions which improve the entanglement detection given by the linear ones are presented in \cite{RefJ:Arrazola} with explicit examples showing accessible nonlinear \emph{EW}s detect more states than their linear ancestors.
\par
The other way for constructing nonlinear \emph{EW}s is based on the \emph{PPT} entangled states detection by improving the linear \emph{EW}s. In this approach for a given density matrix the \emph{PPT} convex region is determined by the \emph{PPT} inequalities of the density matrix. For some \emph{PPT} states in this region, which called the feasible region, the \emph{PPT} criterion is sufficient for separability. Then linear \emph{EW}s introduce. Then nonlinear \emph{EW}s has been constructed from linear ones. This method has been applied for three qubits \emph{MUB} diagonal entangled states in \cite{RefJ:Jafarizadeh_1},  for $ 2\otimes 2 \otimes d $ bound entangled density matrices by exact convex optimization in \cite{RefJ:Jafarizadeh_2}, for general algorithm for manipulating nonlinear and linear entanglement witnesses by using exact convex optimization \cite{RefJ:Jafarizadeh_3}, and for bipartite $ N \otimes N $ systems via exact convex optimization in \cite{RefJ:Jafarizadeh_4}.
\par
In this paper for a given four qubits Hamiltonian or density matrix we determine the $ EW $s with the ability to detect the $ PPT $ entangled states. First, we specify the $ PPT $ region for a given four qubits density matrix in the mutually unbiased basis $ (MUB) $. This region forms a convex region called the feasible region (\emph{FR}). Then we introduce the linear \emph{EW}s family which can detect the \emph{MUB} diagonal density matrices with positive partial transposes. Then we construct the nonlinear \emph{EW}s with the nonlinear coefficients which have wider range detection. These nonlinear \emph{EW}s are envelope the family of previous linear \emph{EW}s and to support the idea, we present an example with full details. This framework helps to investigate the \emph{EW}s for a given density matrix (here four qubits) and serves four in organizing the knowledge about the entanglement of the system. In the last section we study thermal entanglement for an ensemble of four qubits systems in equilibrium. The result shows for special cases our nonlinear witnesses can detect the entanglement at any temperature for some coupling constants.

\section{MUB diagonal density matrices and positive partial transpose conditions }\label{section:2}
Here we review the $ MUB $ bases then we consider a diagonal Hamiltonian and corresponding diagonal density matrix at this base for a system with four qubits. After representing the density matrix in the Pauli matrices bases, we will find the $ PPT $ region explicitly.
\par
Mutually unbiased bases ($ MUB $) in $ N $ dimensional Hilbert space are orthonormal bases $ {|v_{i} \rangle} $ and $ {|w_{j} \rangle} $ such that $ |\langle v_{i} | w_{j}\rangle| = 1/\sqrt{N} $ for all $ i,\ j \in \{ 1,..,N \} $. If one can find $ N + 1 $ mutually unbiased bases for a complex vector space of $ N $ dimensions, then the measurements corresponding to these bases provide an optimal means of determining the density matrix of an ensemble of systems \cite{RefJ:Wootters,RefJ:Durt,RefB:Klappenecker}. These bases may be used for entanglement detection \cite{RefJ:Spengler,RefJ:Chen,RefJ:Ma}.
\par
The Bell basis is an orthonormal basis for the two qubits Hilbert space and in terms of computational basis could be written as
\begin{eqnarray}
|\psi^{-}\rangle &=& \frac{1}{\sqrt{2}}( |01\rangle\ - |10\rangle), \ 
|\psi^{+}\rangle = \frac{1}{\sqrt{2}}( |01\rangle\ + |10\rangle), \nonumber\\    
|\phi^{-}\rangle &=& \frac{1}{\sqrt{2}}( |00\rangle\ - |11\rangle), \
|\phi^{+}\rangle = \frac{1}{\sqrt{2}}( |00\rangle\ + |11\rangle).
\end{eqnarray}
Bell state is any quantum state in the Bell basis. Density matrices which are diagonal in this basis are called Bell-diagonal. In the case of two qubit the Bell-diagonal state is
\begin{equation}
\rho_{_{2\otimes 2}} = 
p_1 |\phi^{+}\rangle \langle \phi^{+}| + 
p_2 |\psi^{+}\rangle \langle \psi^{+}| + 
p_3 |\psi^{-}\rangle \langle \psi^{-}| + 
p_4 |\phi^{-}\rangle \langle \phi^{-}| 
\end{equation}    
where $ 0 \leq p_i \leq 1 $ and $ \sum_{i=1}^{4} p_{i} =1 $. 
\par
The Bell basis can be generalized, specifically consider a system of four qubits spins, the generalized $ 16 $ elements can be written as
\begin{eqnarray}
|\psi_{_{_{1}}}\rangle=\frac{1}{\sqrt{2}}( |0000 \rangle\!+\! |1111\rangle),\;
|\psi_{_{_{2}}}\rangle=\frac{1}{\sqrt{2}}( |0000\rangle\!-\!|1111\rangle )
\nonumber\\
|\psi_{_{_{3}}}\rangle=\frac{1}{\sqrt{2}}( |0001\rangle\!+\!|1110\rangle ),\; |\psi_{_{_{4}}}\rangle=\frac{1}{\sqrt{2}}( |0001\rangle\!-\!|1110\rangle )
\nonumber\\
|\psi_{_{_{5}}}\rangle=\frac{1}{\sqrt{2}}( |0010\rangle\!+\!|1101\rangle ),\; |\psi_{_{_{6}}}\rangle=\frac{1}{\sqrt{2}}( |0010\rangle\!-\!|1101\rangle )
\nonumber\\
|\psi_{_{_{7}}}\,\,\rangle=\frac{1}{\sqrt{2}}( |0011\rangle\!+\!|1100\rangle ),\; |\psi_{_{_{8}}}\,\,\rangle=\frac{1}{\sqrt{2}}( |0011\rangle\!-\!|1100\rangle )
\nonumber\\
|\psi_{_{_{9}}}\:\,\rangle=\frac{1}{\sqrt{2}}( |0100\rangle\!+\!|1011\rangle ),\; |\psi_{_{_{10}}}\rangle=\frac{1}{\sqrt{2}}( |0100\rangle\!-\!|1011\rangle )
\nonumber\\
|\psi_{_{_{11}}}\rangle=\frac{1}{\sqrt{2}}(|0101\rangle\!+\!|1010\rangle ),\; |\psi_{_{_{12}}}\rangle=\frac{1}{\sqrt{2}}( | 0101\rangle\!-\!|1010\rangle )
\nonumber\\
|\psi_{_{_{13}}}\rangle=\frac{1}{\sqrt{2}}(|0110\rangle\!+\!|1001\rangle ),\; |\psi_{_{_{14}}}\rangle=\frac{1}{\sqrt{2}}( | 0110\rangle\!-\!|1001\rangle )
\nonumber\\
|\psi_{_{_{15}}}\rangle=\frac{1}{\sqrt{2}}(|0111\rangle\!+\!|1000\rangle ),\; |\psi_{_{_{16}}}\rangle=\frac{1}{\sqrt{2}}( | 0111\rangle\!-\!|1000\rangle)
\nonumber
\end{eqnarray}
(Other bases choices are possible, for example, see \cite{RefJ:Jaeger_1,RefJ:Jaeger_2}). The diagonal Hamiltonian in these bases is
\begin{equation}\label{Diagonal_Hamiltonian}
H = \sum_{i=1}^{16} E_{i} | \psi_{_{i}} \rangle \langle \psi_{_{i}} |
\end{equation}
where $ E_{i} $ is the energy eigenvalue of the $ | \psi_{_{i}} \rangle $ state. In terms of Pauli spin matrices 
\begin{eqnarray}\label{Hamiltonian_Pauli_Bases}
H & = &c_{_{0}}IIII + c_{_{1}} I\sigma_z\sigma_zI + c_{_{2}} I\sigma_zI\sigma_z + c_{_{3}} II\sigma_z\sigma_z + 
c_{_{4}}\sigma_z I I \sigma_z+ 
\nonumber\\&&
c_{_{5}}\sigma_zI\sigma_zI + 
c_{_{6}}\sigma_z\sigma_zII+
c_{_{7}}\sigma_z\sigma_z\sigma_z\sigma_z +
c_{_{8}}\sigma_x\sigma_x\sigma_x\sigma_x+
\nonumber\\&&
c_{_{9}}\sigma_x\sigma_y\sigma_y\sigma_x +
c_{_{_{10}}}\sigma_x\sigma_y\sigma_x\sigma_y +
c_{_{_{11}}}\sigma_x\sigma_x\sigma_y\sigma_y + c_{_{_{12}}}\sigma_y\sigma_x\sigma_y\sigma_x +
\nonumber\\&&
c_{_{_{13}}}\sigma_y\sigma_x\sigma_x\sigma_y +
c_{_{_{14}}}\sigma_y\sigma_y\sigma_x\sigma_x + c_{_{_{15}}}\sigma_y\sigma_y\sigma_y\sigma_y
\end{eqnarray}
( the tensor product sign is omitted for simplicity, for example the fourth term $ \sigma_z I I \sigma_z $ means $ \sigma_z \otimes I \otimes I \otimes \sigma_z $ ) here $ c_{i} $s can be driven in terms of $ E_{i} $s and characterize the coupling strength among qubits. The first term represents no interaction at all (a constant term), the next six terms represent pair $ z $ component spin interaction (Ising like), and the remaining terms represent the four party interactions. 
\par
Now suppose we have a large number (theoretically, infinite) of four qubits molecules in thermodynamic equilibrium (canonical ensemble). If we assume that the inter-molecular interactions are negligible, then the total system is in a product state, $ \rho \otimes \dots \otimes \rho $, it follows from the additive property of entanglement that the total entanglement present in the system is $ N $ times the entanglement present in a single molecule, where $ N $ is the total number of molecules present in the system \cite{RefT:Nielsen}.
\par
For a canonical ensemble of four qubits in the thermal equilibrium the state of this system in the Bell-diagonal bases can be written as
\begin{equation}\label{FourQubitBellDiagonalDensityMatrix}
\rho=\sum_{i=1}^{16} p_{i} \
|\psi_{i}\rangle
\langle \psi_{i}|
\end{equation}
here 
\begin{equation}\label{pis}
p_{i} = \frac{e^{-\beta E_{i}}}{\sum_{j=1}^{16}e^{-\beta E_{i}}}
\end{equation}
is the probability of finding the system in the state $ |\psi_{i}\rangle $, and $ \beta = \frac{1}{k_{B} T}$, $ k_{B} $ is the Boltzmann constant, $ T $ is the temperature and $\sum_{i=1}^{16} p_{i}=1 $.
\par
The density matrix in terms of two dimensional Pauli matrices is presented in Appendix. We interested to find the positive partial transposition (\emph{PPT}) region. If we consider the following notation
\begin{equation}\label{pi}
\textbf{Set} (p_{_{i}},p_{_{j}},p_{_{k}},p_{_{l}}):=
\left\{\begin{array}{r}
p_{_{i}} + p_{_{j}} + p_{_{k}}-p_{_{l}}\geq0\\
p_{_{i}} + p_{_{j}} - p_{_{k}}+p_{_{l}}\geq0\\
p_{_{i}} - p_{_{j}} + p_{_{k}}+p_{_{l}}\geq0\\
-p_{_{i}} + p_{_{j}} +p_{_{k}}+p_{_{l}}\geq0
\end{array}\right.
\end{equation}
then the positivity conditions for the eigenvalues of $ \rho^{T_A} $ are
\begin{equation}\label{PPT_TA}
\left\{\begin{array}{l}
\textbf{Set} (p_{_{1}},p_{_{2}},p_{_{15}},p_{_{16}})\\
\textbf{Set} (p_{_{3}},p_{_{4}},p_{_{13}},p_{_{14}})\\
\textbf{Set} (p_{_{5}},p_{_{6}},p_{_{11}},p_{_{12}})\\
\textbf{Set} (p_{_{7}},p_{_{8}},p_{_{9}},p_{_{10}})
\end{array}\right.
\end{equation}
for the eigenvalues of $\rho^{T_B}$
\begin{equation}\label{PPT_TB}
\left\{\begin{array}{l}
\textbf{Set} (p_{_{1}},p_{_{2}},p_{_{9}},p_{_{10}})\\
\textbf{Set} (p_{_{3}},p_{_{4}},p_{_{11}},p_{_{12}})\\
\textbf{Set} (p_{_{5}},p_{_{6}},p_{_{13}},p_{_{14}})\\
\textbf{Set} (p_{_{7}},p_{_{8}},p_{_{15}},p_{_{16}})
\end{array}\right.
\end{equation}
for the eigenvalues of $\rho^{T_C}$
\begin{equation}\label{PPT_TC}
\left\{\begin{array}{l}
\textbf{Set} (p_{_{1}},p_{_{2}},p_{_{5}},p_{_{6}})\\
\textbf{Set} (p_{_{3}},p_{_{4}},p_{_{7}},p_{_{8}})\\
\textbf{Set} (p_{_{9}},p_{_{10}},p_{_{13}},p_{_{14}})\\
\textbf{Set} (p_{_{11}},p_{_{12}},p_{_{15}},p_{_{16}})
\end{array}\right.
\end{equation}
for the eigenvalues of $\rho^{T_D}$
\begin{equation}\label{PPT_TD}
\left\{\begin{array}{l}
\textbf{Set} (p_{_{1}},p_{_{2}},p_{_{3}},p_{_{4}})\\
\textbf{Set} (p_{_{5}},p_{_{6}},p_{_{7}},p_{_{8}})\\
\textbf{Set} (p_{_{9}},p_{_{10}},p_{_{11}},p_{_{12}})\\
\textbf{Set} (p_{_{13}},p_{_{14}},p_{_{15}},p_{_{16}})
\end{array}\right.
\end{equation}
for the eigenvalues of $\rho^{T_{AB}}$
\begin{equation}\label{PPT_TAB}
\left\{\begin{array}{l}
\textbf{Set} (p_{_{1}},p_{_{2}},p_{_{7}},p_{_{8}})\\
\textbf{Set} (p_{_{3}},p_{_{4}},p_{_{5}},p_{_{6}})\\
\textbf{Set} (p_{_{9}},p_{_{10}},p_{_{15}},p_{_{16}})\\
\textbf{Set} (p_{_{11}},p_{_{12}},p_{_{13}},p_{_{14}})
\end{array}\right.
\end{equation}
for the eigenvalues of $\rho^{T_{AC}}$
\begin{equation}\label{PPT_TAC}
\left\{\begin{array}{l}
\textbf{Set} (p_{_{1}},p_{_{2}},p_{_{11}},p_{_{12}})\\
\textbf{Set} (p_{_{3}},p_{_{4}},p_{_{9}},p_{_{10}})\\
\textbf{Set} (p_{_{5}},p_{_{6}},p_{_{15}},p_{_{16}})\\
\textbf{Set} (p_{_{7}},p_{_{8}},p_{_{13}},p_{_{14}})
\end{array}\right.
\end{equation}
and finally for the eigenvalues of $\rho^{T_{AD}}$
\begin{equation}\label{PPT_TAD}
\left\{\begin{array}{l}
\textbf{Set} (p_{_{1}},p_{_{2}},p_{_{13}},p_{_{14}})\\
\textbf{Set} (p_{_{3}},p_{_{4}},p_{_{15}},p_{_{16}})\\
\textbf{Set} (p_{_{5}},p_{_{6}},p_{_{9}},p_{_{10}})\\
\textbf{Set} (p_{_{7}},p_{_{8}},p_{_{11}},p_{_{12}})
\end{array}\right.
\end{equation}
The total set of the above $ 4 \times 4 \times 7 = 112 $ inequalities, defines the \emph{PPT} region of the four qubits Bell diagonal states (\ref{FourQubitBellDiagonalDensityMatrix}).

\subsection{ Feasible regions }\label{subsection:2.1}
The $PPT$ conditions, (\ref{PPT_TA}),...,(\ref{PPT_TAD}) are linear inequalities. Each of these inequalities determines a certain half-space while all the inequalities together determine a certain region in $16$-dimensional space, ($p_1,...,p_{16} $). This region, which is the intersection of $112$ half-spaces is called a convex polyhedral region, which is the feasible region of the problem \cite{RefB:Solodovnikov}.
\par
As this $FR$ is complex for detail study, we can consider the eight partitions of the $ 112 $ inequalities as follows
$$
(p_{_{1}},p_{_{2}}),\
(p_{_{3}},p_{_{4}}),\
(p_{_{5}},p_{_{6}}),\
(p_{_{7}},p_{_{8}}),\
(p_{_{9}},p_{_{10}}),\
(p_{_{11}},p_{_{12}}),\
(p_{_{13}},p_{_{14}}),\
(p_{_{15}},p_{_{16}})
$$
here each partition means the inequalities is contained to relative $p_i$'s. For example, $ (p_{_{1}},p_{_{2}}) $, means the all inequalities which contain $p_{_{1}}$ and $p_{_{2}}$. Now if we choose one pair such as $(p_{_{1}},p_{_{2}})$, then we can specify the feasible region in $(p_{_{1}},p_{_{2}})$ plane  with the following three inequalities (\emph{PPT} conditions)
$$
\left\{\begin{array}{lcr}
p_{_{1}}&\leq& p_{_{2}}+\;p_{_{3}}+\;p_{_{4}}\\
p_{_{1}}&\leq& p_{_{2}}+\;p_{_{5}}+\;p_{_{6}}\\
p_{_{1}}&\leq& p_{_{2}}+\;p_{_{7}}+\;p_{_{8}}\\
p_{_{1}}&\leq& p_{_{2}}+\;p_{_{9}}+p_{_{10}}\\
p_{_{1}}&\leq& p_{_{2}}+p_{_{11}}+p_{_{12}}\\
p_{_{1}}&\leq& p_{_{2}}+p_{_{13}}+p_{_{14}}\\
p_{_{1}}&\leq& p_{_{2}}+p_{_{15}}+p_{_{16}}
\end{array}\right.
$$
Adding both sides of above inequalities together and noting $ p_{_1}+...+p_{_{16}}=1 $, yields the following inequality
\begin{eqnarray}\label{p1p2Inequality_1}
8p_{_{1}}-6p_{_{2}}\leq1
\end{eqnarray}
similarly,
\begin{eqnarray}\label{p1p2Inequality_2}
8p_{_{2}}-6p_{_{1}}\leq1
\end{eqnarray} 

\begin{figure}
	\centering
	\includegraphics[width=0.7\textwidth]{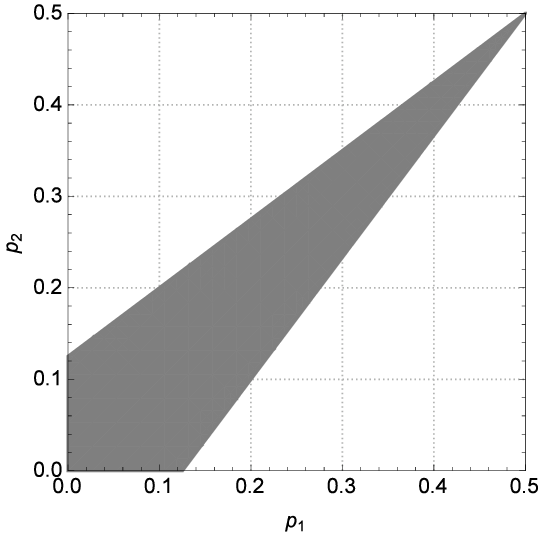}
	\caption{The PPT feasible region for four qubits Bell diagonal states. For all points in the shadow region all eigenvalues of the density matrix (\ref{FourQubitBellDiagonalDensityMatrix}), and all partial transposes are positive. The region is for inequalities (\ref{p1p2Inequality_1}) and (\ref{p1p2Inequality_2}).}
	\label{Fig1_p1_p2_PPT_Feasible_Region}      
\end{figure}

This region is illustrated in Fig.~\ref{Fig1_p1_p2_PPT_Feasible_Region}. This region is convex and we show each vertex in this region satisfies the all $PPT$ conditions ($112$ inequalities), that is to say, each vertex is inside of the $FR$ or polyhedron. To do this consider the vertex $(p_1=0, p_2=0)$. After setting $(p_1=0, p_2=0)$ in the $112$ inequalities, we obtain (see Sec.~\ref{subsection:2.2})
\begin{eqnarray}\label{p1=0,p2=0}
p_{_{3}} = p_{_{4}},\ p_{_{5}} = p_{_{6}},\ p_{_{7}} = p_{_{8}},\ p_{_{9}} = p_{_{10}},\ p_{_{11}}= p_{_{12}},\ p_{_{13}}= p_{_{14}},\ p_{_{15}}= p_{_{16}}
\nonumber\\
\end{eqnarray}
As these equations have solutions, ( for example $p_{_{3}}=\cdots =p_{_{16}} =1/14$), then the vertex $(p_1=0,\ p_2=0)$ lies inside of the $FR$. For the second vertex, $(p_1=1/8,\ p_2=0)$, we obtain
\begin{eqnarray}\label{p1=1/8,p2=0}
p_{_{3}} + p_{_{4}}\geq 1/8,\ p_{_{5}} + p_{_{6}}\geq 1/8,\ p_{_{7}} + p_{_{8}}\geq 1/8,\nonumber\\
p_{_{9}} + p_{_{10}}\geq 1/8,\ p_{_{11}}+ p_{_{12}}\geq 1/8,\ p_{_{13}} + p_{_{14}}\geq 1/8,\nonumber\\
p_{_{15}} + p_{_{16}}\geq 1/8,\ -1/8 \leq p_{_{13}} - p_{_{14}} \leq 1/8,\nonumber\\
\textbf{Set} (p_{_{3}},p_{_{4}},p_{_{15}},p_{_{16}}),\
\textbf{Set} (p_{_{5}},p_{_{6}},p_{_{9}},p_{_{10}}),\
\textbf{Set} (p_{_{7}},p_{_{8}},p_{_{11}},p_{_{12}})
\nonumber\\
\end{eqnarray}
These sets of inequalities have solution such as $p_{_{3}}=\cdots =p_ {_ {16}} =1/16$, so this vertex belongs to the $FR$. The same argument is valid for third vertex $(p_1=0,\ p_2=1/8)$. Finally, for the fourth vertex $(p_1=1/2,\ p_2=1/2) $ the $PPT$ condition inequalities reduce to 
\begin{equation}
p_{_{3}}=\cdots =p_{_{16}} =0
\end{equation}
so all vertexes belong to the $FR$ and satisfy the all $PPT$ conditions. As $(p_{_{1}},\ p_{_{2}})$ region in Fig.~\ref{Fig1_p1_p2_PPT_Feasible_Region} is convex, then the all points of this region are inside the total $FR$.
\par
We can also find other feasible regions in other planes such as $(p_{_{1}} ,p_{_{3}})$ plane, concerning the following  inequalities
\begin{equation}\label{inequ_p1p3p2p4}
\left\{\begin{array}{lcr}
p_{_{1}}&\leq& p_{_{2}}+\;\, p_{_{5}}+\; p_{_{6}}\\
p_{_{1}}&\leq& p_{_{2}}+\;\, p_{_{7}}+\; p_{_{8}}\\
p_{_{1}}&\leq& p_{_{2}}+\; p_{_{9}}+p_{_{10}}\\
p_{_{3}}&\leq& p_{_{4}}+p_{_{11}}+p_{_{12}}\\
p_{_{3}}&\leq& p_{_{4}}+p_{_{13}}+p_{_{14}}\\
p_{_{3}}&\leq& p_{_{4}}+p_{_{15}}+p_{_{16}}\\
\end{array}\right.\Rightarrow 
4(p_{_{1}}+p_{_{3}})\leq 2(p_{_{2}}+p_{_{4}})+1
\end{equation}
similarly we obtain
\begin{equation}\label{inequ_p2p4p1p3}
4(p_{_{2}}+p_{_{4}})\leq 2(p_{_{1}}+p_{_{3}})+1
\end{equation}
from inequalities (\ref{inequ_p1p3p2p4}) and (\ref{inequ_p2p4p1p3}), we have
\begin{equation}\label{inequ_p1p3}
4(p_{_{1}}+p_{_{3}})\leq 2(p_{_{2}}+p_{_{4}})+1 \leq 
p_{_{1}}+p_{_{3}}+\frac{3}{2} 
\end{equation}
or
\begin{equation}
p_{_{1}}+p_{_{3}}\leq \frac{1}{2}
\end{equation}
so we have presented a new perspective from the spatial shape in two-dimensions. There are many such perspectives which the reader can investigate using the $PPT$ inequalities. We present another perspectives of the feasible region in two cases in the Appendix.

\subsection{\emph{MUB} diagonal states which the \emph{PPT} criterion is necessary and sufficient condition for separability}\label{subsection:2.2}
In this section we investigate some \emph{MUB} diagonal states which the \emph{PPT} criterion is the necessary and sufficient condition for the separability of them. To this end, we write $p_{i}$ in the following pairs
\begin{eqnarray}
(p_{_{1}},p_{_{2}}),\  (p_{_{3}},p_{_{4}}),\ (p_{_{5}},p_{_{6}}),\ (p_{_{7}},p_{_{8}}),\  (p_{_{9}},p_{_{10}}),\ (p_{_{11}},p_{_{12}}),\ (p_{_{13}},p_{_{14}}),\ (p_{_{15}},p_{_{16}})
\end{eqnarray}
Note that when any pair is zero then the two components of others are equal and the \emph{PPT} criterion is necessary and sufficient for separability of the \emph{MUB} diagonal density matrix. For example, if we set $p_{_{1}}=p_{_{2}}=0$ in the first pair, then from first set of(\ref{PPT_TD}), we have
\begin{equation}\label{pi}
\textbf{Set} (p_{_{1}}=0,p_{_{2}}=0,p_{_{3}},p_{_{4}}):=
\left\{\begin{array}{r}
p_{_{3}}+p_{_{4}}\geq0\\
- p_{_{3}}+p_{_{4}}\geq0\\
p_{_{3}} - p_{_{4}} \geq0
\end{array}\right.
\end{equation}
or $p_{_{4}} \geq p_{_{3}},\ p_{_{3}} \geq p_{_{4}}$ which is equal with $p_{_{3}}=p_{_{4}}$. Similarly we can show that
$p_{_{5}}=p_{_{6}} ,p_{_{7}}=p_{_{8}} ,p_{_{9}}=p_{_{10}} ,p_{_{11}}=p_{_{12}},p_{_{13}}=p_{_{14}}$ and $p_{_{15}}=p_{_{16}}.$ Now we can write the MUB diagonal density matrix in the following form
\begin{eqnarray}
\rho& = &\frac{1}{8} \Big[
2 p_{_{3}}(|\psi_{_{3}}\rangle \langle \psi_{_{3}}|\!+\!|\psi_{_{4}}\rangle \langle \psi_{_{4}}|)\!+\!2 p_{_{5}} (|\psi_{_{5}}\rangle \langle \psi_{_{5}}|\!+\!|\psi_{_{6}}\rangle \langle \psi_{_{6}}|)\nonumber\\&&
+ 2 p_{_{7}}(|\psi_{_{7}}\rangle \langle \psi_{_{7}}|\!+\!|\psi_{_{8}}\rangle \langle \psi_{_{8}}|)
+ 2p_{_{9}}(|\psi_{_{9}}\rangle \langle \psi_{_{9}}|\!+\!|\psi_{_{10}}\rangle \langle \psi_{_{10}}|) \nonumber\\&&
+ 2 p_{_{11}}(|\psi_{_{11}}\rangle \langle \psi_{_{11}} |\!+\!
|\psi_{_{12}}\rangle \langle \psi_{_{12}}|)\nonumber\\&&
+ 2p_{_{13}}(|\psi_{_{13}}\rangle \langle \psi_{_{13}}|\!+\!
|\psi_{_{14}}\rangle \langle \psi_{_{14}}| )
+2p_{_{15}}(|\psi_{_{15}}\rangle \langle \psi_{_{15}}|\!+\!
|\psi_{_{16}}\rangle \langle \psi_{_{16}}|)
\Big]
\end{eqnarray}
which is a separable state.

\section{Witnesses detecting bound \emph{MUB} diagonal density matrices}\label{section:3}
We introduce our linear four qubits entanglement witnesses that have the following generic form
\begin{eqnarray}\label{NonLinearEWsGenericForm}
W & = & A_0IIII\pm B_0 \sigma_z\sigma_zII+\nonumber\\&&
A_{_{1}}(\sigma_x\sigma_x\sigma_x\sigma_x+\sigma_x\sigma_x\sigma_y\sigma_y)+
A_{_{2}}(\sigma_y\sigma_y\sigma_x\sigma_x+\sigma_y\sigma_y\sigma_y\sigma_y)+\nonumber\\&&
A_{_{3}}(\sigma_x\sigma_y\sigma_y\sigma_x+\sigma_x\sigma_y\sigma_x\sigma_y)+
A_{_{4}}(\sigma_y\sigma_x\sigma_y\sigma_x+\sigma_y\sigma_x\sigma_x\sigma_y)
\end{eqnarray}
In order to investigate whether this operator really is an entanglement witness we must first prove its expectation value over separable states is nonnegative. To do so we evaluate the trace of witness over a pure product state \footnote{As any separable state can be written as a convex combination of pure product states, namely $ \rho_s = \sum_{i} h_{i} |\nu_{i} \rangle \langle \nu_{i} |,\ 0\leq h_{i} \leq1,\ \sum_{i} h_{i} =1 $, then it is sufficient to follow the proof for one product state.} which for four qubits state may be written as
$
\rho_s=|\alpha \rangle \langle \alpha |\otimes |\beta \rangle \langle \ \beta |\otimes | \gamma \rangle \langle \gamma|\otimes |\lambda \rangle \langle \ \lambda|
$. The trace takes the following form
\begin{eqnarray}
Tr(W \rho_s) & = &
A_0 \pm B_0 a_{_{3}}b_{_{3}} + \nonumber\\&&
A_{_{1}}(a_{_{1}}b_{_{1}}c_{_{1}}d_{_{1}} + a_{_{1}}b_{_{1}}c_{_{2}}d_{_{2}}) +
A_{_{2}}(a_{_{2}}b_{_{2}}c_{_{1}}d_{_{1}} + a_{_{2}}b_{_{2}}c_{_{2}}d_{_{2}}) +\nonumber\\&&
A_{_{3}}(a_{_{1}}b_{_{2}}c_{_{2}}d_{_{1}} + a_{_{1}}b_{_{2}}c_{_{1}}d_{_{2}}) + A_{_{4}}(a_{_{2}}b_{_{1}}c_{_{2}}d_{_{1}} + a_{_{2}}b_{_{1}}c_{_{1}}d_{_{2}})
\end{eqnarray}
where
\[ \begin{array}{l}
Tr(|\alpha\rangle\langle\alpha|\sigma_i) = a_i , \  
Tr(|\beta\rangle\langle\beta|\sigma_i) = b_i , \\
Tr(|\gamma\rangle\langle\gamma|\sigma_i) = c_i , \ 
Tr(|\lambda\rangle\langle\lambda|\sigma_i) = d_i 
\end{array}\]
for $ i= 1, 2, 3 $ and  $\sigma_i $'s are spin $ 1/2 $ Pauli matrices. With definitions
\[ \begin{array}{lclcr}
a_{_{1}}=sin{\theta_{_{1}}}cos{\varphi_{_{1}}}&, & a_{_{2}}=sin{\theta_{_{1}}}sin{\varphi_{_{1}}}&, &
a_{_{3}}=cos{\theta_{_{1}}}\\
b_{_{1}}=sin{\theta_{_{2}}}cos{\varphi_{_{2}}}&, & b_{_{2}}=sin{\theta_{_{2}}}sin{\varphi_{_{2}}}&, &
b_{_{3}}=cos{\theta_{_{2}}}\\
c_{_{1}}=sin{\theta_{_{3}}}cos{\varphi_{_{3}}}&, &
c_{_{2}}=sin{\theta_{_{3}}}sin{\varphi_{_{3}}}&, &
c_{_{3}}=cos{\theta_{_{3}}}\\
d_{_{1}}=sin{\theta_{_{4}}}cos{\varphi_{_{4}}}&, &
d_{_{2}}=sin{\theta_{_{4}}}sin{\varphi_{_{4}}}&, &
d_{_{3}}=cos{\theta_{_{4}}}
\end{array}\]
the $ Tr(W\rho_s) $, takes the following simple form
\begin{eqnarray}
Tr(W\rho_s)&=& A_0\pm B_0\cos\theta_{_{1}}\cos\theta_{_{2}}+ \nonumber\\&& \sin\theta_{_{1}}\sin\theta_{_{2}}\sin\theta_{_{3}}\sin\theta_{_{4}}
\bigg{\{ } \nonumber\\&& 
\cos(\varphi_{_{3}}-\varphi_{_{4}})
\big{(} A_1 \cos\varphi_{_{1}}\cos\varphi_{_{2}}+A_2 \sin\varphi_{_{1}}\sin\varphi_{_{2}}\big{)}+\nonumber\\&&
\sin(\varphi_{_{3}}+\varphi_{_{4}})
\big{(}A_3 \cos\varphi_{_{1}}\sin\varphi_{_{2}}+A_4 \cos\varphi_{_{2}}\sin\varphi_{_{1}}\big{)}
\bigg{\}}
\end{eqnarray}
If we define new parameters
\begin{eqnarray}
& &
h_{_{1}} = \frac{A_{_{1}} + A_{_{2}}}{2}, \;
h_{_{2}} = \frac{A_{_{1}} - A_{_{2}}}{2}, \;
h_{_{3}} = \frac{A_{_{3}} + A_{_{4}}}{2}, \;
h_{_{4}} = \frac{A_{_{3}} - A_{_{4}}}{2}
\end{eqnarray}
then
\begin{eqnarray}
Tr(W\rho_s)&=& A_0\pm B_0\cos\theta_{_{1}}\cos\theta_{_{2}}+ \nonumber\\&& \sin\theta_{_{1}}\sin\theta_{_{2}}\sin\theta_{_{3}}\sin\theta_{_{4}}
\bigg{\{} \nonumber\\&&
\cos(\varphi_{_{3}}-\varphi_{_{4}})
\bigg[ h_{_{1}}\cos(\varphi_{_{1}}-\varphi_{_{2}}) + h_{_{2}}\cos(\varphi_{_{1}} + \varphi_{_{2}})\bigg]+
\nonumber\\& &
\sin(\varphi_{_{3}} + \varphi_{_{4}})
\bigg[ h_{_{3}}\sin(\varphi_{_{1}} + \varphi_{_{2}}) + h_{_{4}}\sin(\varphi_{_{2}} - \varphi_{_{1}})\bigg] \bigg{\}}
\nonumber
\end{eqnarray}
By appropriate choice of the angles, one can minimize  above expression, where its minimum value must be zero. For this purpose, we set $\theta_{_{3}} = \theta_{_{4}} = \frac{\pi}{2} $, $ \varphi_{_{3}} = \varphi_{_{4}} = \frac{\pi}{4} $, and define new parameters
\begin{eqnarray}
& &
\cos \psi_{1} = \frac{h_{1}}{\sqrt{h_{1}^{2}+h_{4}^{2}}}, \;
\cos \psi_{2} = \frac{h_{2}}{\sqrt{h_{2}^{2}+h_{3}^{2}}}, \;
\sin \psi_{1} = \frac{h_{4}}{\sqrt{h_{1}^{2}+h_{4}^{2}}}, \;
\sin \psi_{2} = \frac{h_{3}}{\sqrt{h_{2}^{2}+h_{3}^{2}}}
\nonumber
\end{eqnarray}
then
\begin{eqnarray}
Tr( W \rho_s )&=&
A_0 \pm B_0\cos\theta_{_{1}}\cos\theta_{_{2}} +
\sin\theta_{_{1}}\sin\theta_{_{2}} \Big\{
\nonumber\\& &
\sqrt{h_{1}^{2}+h_{4}^{2}} \left[  \cos\psi_{1} \cos(\varphi_{1}-\varphi_{2})+\sin\psi_{1} \sin(\varphi_{1}+\varphi_{2}) 
\right]
\nonumber\\& &
\sqrt{h_{2}^{2}+h_{3}^{2}} \left[  \cos\psi_{2} \cos(\varphi_{1}+\varphi_{2})+\sin\psi_{2} \sin(\varphi_{1}+\varphi_{2}) 
\right]  \Big\}
\nonumber\\& &
=A_0 \pm B_0\cos\theta_{_{1}}\cos\theta_{_{2}} +
\sin\theta_{_{1}}\sin\theta_{_{2}} \Big\{
\nonumber\\& &
\sqrt{h_{1}^{2}+h_{4}^{2}}   \cos(\psi_{1}-\varphi_{1}+\varphi_{2}) +
\sqrt{h_{2}^{2}+h_{3}^{2}} \cos(\psi_{2}-\varphi_{1}-\varphi_{2})  \Big\}
\nonumber
\end{eqnarray}
Setting $ \psi_{1}=\varphi_{1}-\varphi_{2} $ and $ \psi_{2}=\varphi_{1}+\varphi_{2} $, 
\begin{eqnarray}
Tr( W \rho_s )&=&
A_0 \pm B_0\cos\theta_{_{1}}\cos\theta_{_{2}} +
\sin\theta_{_{1}}\sin\theta_{_{2}} \Big\{
\sqrt{h_{1}^{2}+h_{4}^{2}} +
\sqrt{h_{2}^{2}+h_{3}^{2}} \Big\}
\nonumber
\end{eqnarray}
Using the identity
$$
- \sqrt{\eta^{2}+\delta^{2}} \leq \eta \cos \theta_{2} + \delta \sin \theta_{2} \leq \sqrt{\eta^{2}+\delta^{2}}
$$
where $ \eta $ and $ \delta $ are coefficients of $ \cos \theta_{2} $ and $ \sin \theta_{2} $ respectively, we have
\begin{eqnarray}
Tr( W \rho_s )&\geq&
A_0 \mp
\left[
B_{0}^{2} \cos^{2}\theta_{1}+ \left( \sqrt{h_{1}^{2}+h_{4}^{2}} +
\sqrt{h_{2}^{2}+h_{3}^{2}}\right) ^{2} \sin^{2}\theta_{1}
\right]^{1/2}
\nonumber
\end{eqnarray}
choosing
\begin{equation}
A_0 = B_0 = \sqrt{h_{_{1}}^{2} + h_{_{4}}^{2}} + \sqrt{h_{_{2}}^{2} + h_{_{3}}^{2}}
\end{equation} 
yields to
$$
Tr( W \rho_s ) \geq 0
$$
and the entanglement witness becomes
\begin{eqnarray}
W = A_0 \Big[IIII \pm \sigma_z\sigma_zII \! + \!\frac{A_{_{1}}}{A_0}(\sigma_x\sigma_x\sigma_x\sigma_x\!+\!\sigma_x\sigma_x\sigma_y\sigma_y)+
&&\nonumber\\
\frac{A_{_{2}}}{A_0}(\sigma_y\sigma_y\sigma_x\sigma_x\!+\!\sigma_y\sigma_y\sigma_y\sigma_y)\!+\!
\frac{A_{_{3}}}{A_0}(\sigma_x\sigma_y\sigma_y\sigma_x\!+\!\sigma_x\sigma_y\sigma_x\sigma_y)+
&&\nonumber\\
\!\frac{A_{_{4}}}{A_0}(\sigma_y\sigma_x\sigma_y\sigma_x\!+\!\sigma_y\sigma_x\sigma_x\sigma_y)\Big]
&&\nonumber
\end{eqnarray}
we note that
$$
\frac{A_{_{1}}}{A_0}=\frac{h_{_{1}}+h_{_{2}}}{A_0}=
\frac{\cos\psi_{_{1}}\sqrt{h_{_{1}}^{2}+h_{_{4}}^{2}}+\cos\psi_{_{2}}\sqrt{h_{_{2}}^{2}+h_{_{3}}^{2}}}{A_0}
$$
and if
\begin{equation}
p=\frac{1}{A_0}\sqrt{h_{_{1}}^{2}+h_{_{4}}^{2}} = \frac{1}{2 A_0}\sqrt{2(A_{_{1}}^{2}+A_{_{2}}^{2})}
\end{equation}
then we have
\begin{eqnarray}\label{ps1s2}
&&
\frac{A_{_{1}}}{A_0}=p\cos\psi_{_{1}}+(1-p)\cos\psi_{_{2}}
\nonumber\\&&
\frac{A_{_{2}}}{A_0}=p\cos\psi_{_{1}}-(1-p)\cos\psi_{_{2}}
\\&&
\frac{A_{_{3}}}{A_0}=p\sin\psi_{_{1}}+(1-p)\sin\psi_{_{2}}
\nonumber\\&&
\frac{A_{_{4}}}{A_0}=-p\sin\psi_{_{1}}+(1-p)\sin\psi_{_{2}}
\nonumber
\end{eqnarray}
and the entanglement witness, \emph{W}, can be written as the following form ( without loss of generality we divide the $ W $ by $ A_0 $ )
\begin{eqnarray}\label{Entanglement_Witnesses_Optimal}
W & = & IIII\pm \sigma_z\sigma_z II +
\nonumber\\&&
\big[ p\cos\psi_{_{1}} + (1-p)\cos\psi_{_{2}} \big](\sigma_x\sigma_x\sigma_x\sigma_x + \sigma_x\sigma_x\sigma_y\sigma_y) +
\nonumber\\&&
\big[p\cos\psi_{_{1}}-(1-p)\cos\psi_{_{2}}\big](\sigma_y\sigma_y\sigma_x\sigma_x+\sigma_y\sigma_y\sigma_y\sigma_y) +
\nonumber\\&&
\big[p\sin\psi_{_{1}}+(1-p)\sin\psi_{_{2}}\big](\sigma_x\sigma_y\sigma_y\sigma_x+\sigma_x\sigma_y\sigma_x\sigma_y) +
\nonumber\\&&
\big[\!-p\sin\psi_{_{1}}+(1-p)\sin\psi_{_{2}}\big](\sigma_y\sigma_x\sigma_y\sigma_x+\sigma_y\sigma_x\sigma_x\sigma_y)
\end{eqnarray}
This witness has similar structure as the one from $Eq.~12$ of \cite{RefJ:Guhne_R}, which also needs certain off-diagonal terms. Also in \cite{RefJ:Nagata}, the author shows that any $ N $-qubit state which is diagonal in the $GHZ$ basis is full $ N $-qubit entangled state if and only if no partial transpose of the multi-qubit state is positive with respect to any partition. The reader may be interested to compare the results with these papers.

\section{Nonlinear Entanglement Witnesses}\label{section:5}
We showed that how one can find the \emph{PPT} feasible region and we introduced the linear \emph{EW}s. Now we can construct nonlinear entanglement witnesses for the four qubits \emph{MUB} diagonal states using the envelope definition for a family of curves. 
\begin{figure}\label{fig:Envelope}
	\centering
	\includegraphics[width=0.6\textwidth]{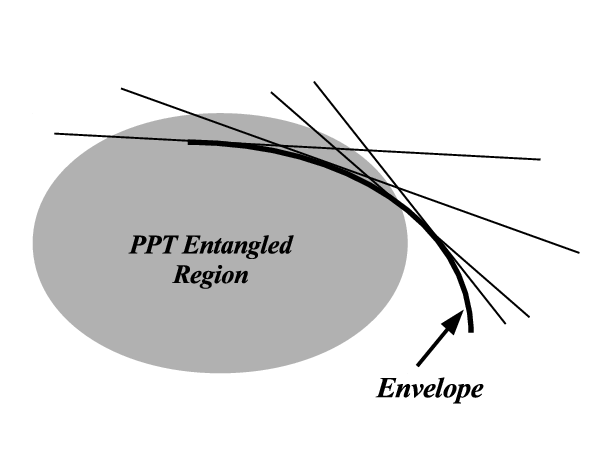}
	\caption{Nonlinear \emph{EW} is the envelope of a family of linear \emph{EW}s. The gray region is the \emph{PPT} entangled or feasible region. Any line represents a possible linear witness. There is some $PPT$ entangled region that cannot be detected by witnesses. The set of linear \emph{EW}s is a family of curves which their envelope can be considered as a nonlinear \emph{EW}.}
	\label{fig:nonlinear}
\end{figure}
\par
Let $ F: \Re \times \Re^{r} \rightarrow \Re $ be a smooth map and $ t, \ x_1,...,\ x_r $ coordinates on the left. Consider \emph{F} as a family of functions $ x $, parameterized by $ t $. The envelope, of the family \emph{F} is the set \cite{RefB:Bruce}
\begin{equation}\label{Envelope_Definition}
\Omega_{F} =\left\lbrace x \in \Re^{r} \textbf{ : there exists } t \in \Re \textbf{ with } F(t,x) = \partial F(t,x) / \partial t =0  \right\rbrace
\end{equation}
Using this definition we can find the envelope of our linear \emph{EW}s. This envelope corresponds to a nonlinear \emph{EW}. To do so, we consider the trace of \emph{EW} over four qubits \emph{MUB} diagonal state, (\ref{FourQubitBellDiagonalDensityMatrix}), as a family of functions (linear \emph{EW}s)
\begin{eqnarray}\label{TrWRoNonlinear}
Tr(W\rho) & = & 1\pm r_{_{6}}+(p\cos\psi_{_{1}}+(1-p)\cos\psi_{_{2}})(r_{_{8}}+r_{_{11}})
\nonumber\\&&
+(p\cos\psi_{_{1}}-(1-p)\cos\psi_{_{2}})(r_{_{14}}+r_{_{15}})
\nonumber\\&&
+(p\sin\psi_{_{1}}+(1-p)\sin\psi_{_{2}})(r_{_{9}}+r_{_{10}})
\nonumber\\&&
+(-p\sin\psi_{_{1}}+(1-p)\sin\psi_{_{2}})(r_{_{12}}+r_{_{13}})
\end{eqnarray}
or in terms of $ p_{i}$ 
\begin{eqnarray}\label{TrWRoNonlinearPs}
Tr(W\rho) & = & 1\pm \left( 1-2 \sum_{j=9}^{16} p_{j} \right)+
\nonumber\\&& 
4 p \left[ (p_{_{_{11}}}\!-p_{_{_{12}}}\!+p_{_{_{13}}}\!-p_{_{_{14}}}) \cos\psi_{1}\!+
(p_{_{_{9}}}\!-p_{_{_{10}}}\!-p_{_{_{15}}}\!+p_{_{_{16}}}) \sin\psi_{1} \right] +
\nonumber\\&&
4 (1-p) \left[
(p_{_{_{3}}}\!-p_{_{_{4}}}\!+p_{_{_{5}}}\!-p_{_{_{6}}}) \cos\psi_{2}\!-
(p_{_{_{1}}}\!-p_{_{_{2}}}\!-p_{_{_{7}}}\!+p_{_{_{8}}}) \sin\psi_{2} \right]
\end{eqnarray}
This family of functions has two parameters, $ \psi_{_{1}} $ and  $ \psi_{_{2}} $ and the condition $ \partial Tr(W \rho) / \partial \psi_{_{1}} =0 $ yields
\begin{equation}\label{EnvelopeCond1}
\psi_{_{1}} =\arctan \left(  \frac{p_{_{_{9}}}\!-p_{_{_{10}}}\!-p_{_{_{15}}}\!+p_{_{_{16}}}}{p_{_{_{11}}}\!-p_{_{_{12}}}\!+p_{_{_{13}}}\!-p_{_{_{14}}}}\right) 
\end{equation}
similarly $ \partial Tr(W \rho) / \partial \psi_{_{2}} =0 $ leads to
\begin{equation}\label{EnvelopeCond2}
\psi_{_{2}} =\arctan \left(  \frac{-p_{_{_{1}}}\!+p_{_{_{2}}}\!+p_{_{_{7}}}\!-p_{_{_{8}}}}{p_{_{_{3}}}\!-p_{_{_{4}}}\!+p_{_{_{5}}}\!-p_{_{_{6}}}}\right)
\end{equation}
now if we insert equations (\ref{EnvelopeCond1}) and (\ref{EnvelopeCond2}) in (\ref{TrWRoNonlinear}) and simplify the result then 
\begin{eqnarray}\label{NonLinearEWsPis}
Tr(W\rho) & = & 
1 + a_{_{0}} \left(1-2\sum_{j=9}^{16} p_{j} \right) 
\nonumber\\&&
+\ 4 p\ a_{_{1}}
\Big[    (p_{_{11}}-p_{_{12}}+p_{_{13}}-p_{_{14}})^{2}+(p_{_{9}}-p_{_{10}}-p_{_{15}}+p_{_{16}})^{2}  \Big]^{1/2} 
\nonumber\\&&
\times\ \textbf{sgn} \
(p_{_{11}}-p_{_{12}}+p_{_{13}}-p_{_{14}}) 
\nonumber\\&&
+\ 4 (1-p)\ a_{_{2}}
\Big[    (p_{_{3}}-p_{_{4}}+p_{_{5}}-p_{_{6}})^{2}+(p_{_{1}}-p_{_{2}}-p_{_{7}}+p_{_{8}})^{2}  \Big]^{1/2}
\nonumber\\&&
\times\ \textbf{sgn} \
(p_{_{3}}-p_{_{4}}+p_{_{5}}-p_{_{6}})
\end{eqnarray}
where $ a_{_{0}} = \pm 1 $, $ \textbf{sgn} $ is the sign function, $ p = \frac{1}{2 A_0}\sqrt{2(A_{_{1}}^{2}+A_{_{2}}^{2})} $, and if $ 2 \pi k - \frac{\pi}{2} \leq \psi_{i} \leq 2 \pi k + \frac{\pi}{2} $ then $ a_{i} $ is $ +1 $, otherwise $ a_{i} $, is $ -1 $ for $ i=1,2 $. Now the envelope equation is $ Tr(W\rho) = 0 $. The following example is given to indicate the full idea and details of the envelope approach for nonlinear witnesses.
\subsection*{\textit{Examples}}
\begin{figure}
	\centering
	\includegraphics[width=0.7\linewidth]{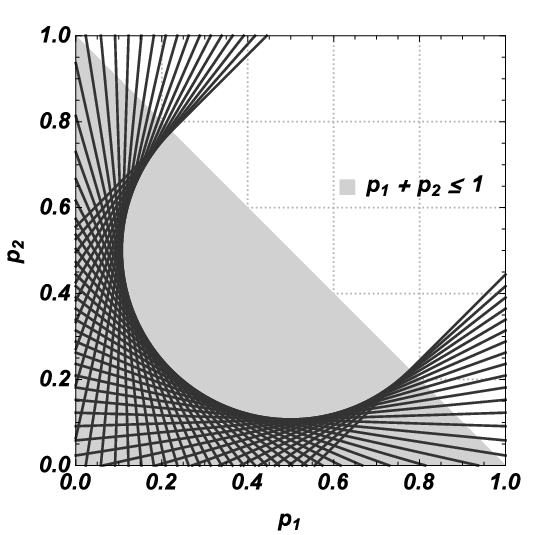}
	\caption[Short Caption]{Linear Entanglement Witnesses. Any point in the shaded region represents a possible four qubits density matrix (not necessarily $PPT$). Each line shows a linear witness separating the entangled density matrices from separable. Here we plotted $50$ linear witnesses for various values of $ \frac{\pi}{2}\leq \psi_{2} \leq \frac{3\pi}{2}$. }
	\label{fig:Fig_3_Example_Linear_Witnesses}
\end{figure}

As the first example we consider the envelope algorithm in operation for a special case where the density matrices are not necessarily $PPT$ and only the nonlinearity of witness as an envelope is investigated. Setting
$$
p =1/10,\ p_{_{i}} =0 \textbf{ for } i\geq 4,\ p_{_{3}}=1-p_{_{1}}-p_{_{2}}
$$
from equation (\ref{TrWRoNonlinearPs}) and taking the plus sign in the second term, we have
\begin{equation}\label{Example_p1p2p3}
Tr(W\rho) = 2 + \frac{18}{5} (p_{2}-p_{1}) \sin \psi_{2}+\frac{18}{5} (1-p_{1}-p_{2}) \cos \psi_{2}
\end{equation}
where $ p_{2},\ p_{2} $ are variables of the density matrix with constraints $ 0 \leq p_{1} \leq 1, \ 0 \leq p_{2} \leq 1,\  0 \leq p_{1}+p_{2} \leq 1 $, and $ 0\leq \psi_{2} \leq 2 \pi $, is a witness parameter. Solving $ Tr(W\rho) = 0 $ in terms of $ p_{2} $ yields to
\begin{equation}\label{Example_p2}
p_{2}=\frac{\cos \psi_{2}-p_{1} (\cos \psi_{2}+\sin \psi_{2})+5/9}{\cos \psi_{2}-\sin \psi_{2}}
\end{equation}
Now any value of $ \psi_{2} $ corresponds to a linear witness.  Fig.~\ref{fig:Fig_3_Example_Linear_Witnesses} shows $ 50 $ of these linear witnesses for $ \psi_{2}=\left\lbrace \frac{\pi}{2},\frac{\pi}{2}+\frac{\pi}{50},\frac{\pi}{2}+\frac{2\pi}{50},\cdots,\frac{3\pi}{2} \right\rbrace  $, in the region restricted to constraints $ 0 \leq p_{1} \leq 1, \ 0 \leq p_{2} \leq 1$, and $ 0 \leq p_{1}+p_{2} \leq 1 $. We can find the envelope equation of these linear witnesses using definition, (\ref{Envelope_Definition}). If we obtain $ \psi_{2} $ from $\partial Tr(W\rho) /\partial \psi_{2} = 0 $ and insert it in Eq.~(\ref{Example_p2}) then
\begin{equation}
p_{2} = \Bigg\{ \begin{array}{l}
\frac{1}{18} \left(9-\sqrt{324 p_{1} (1-p{1})-31}\right),\ \frac{1}{18} \left(9-5 \sqrt{2}\right)\leq p1\leq \frac{7}{9}\\
\frac{1}{18} \left(9+\sqrt{324 p_{1} (1-p{1})-31}\right),\ \frac{1}{18} \left(9-5 \sqrt{2}\right)\leq p1\leq \frac{2}{9}
\end{array}
\end{equation}

\begin{figure}
	\centering
	\includegraphics[width=0.6\linewidth]{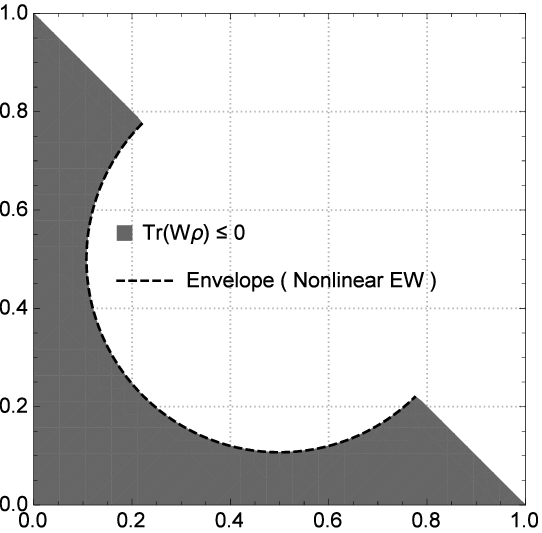}
	\caption{The detected entanglement ( the dark gray region), and nonlinear EW as the envelope ( the dashed curve ) is plotted in special cases which $ a_{_{0}}= +1, \ a_{_{1}}= +1, \ a_{_{2}}= -1,\ p = \frac{1}{10},\ p_{_{i}} =0 \textbf{ for } i\geq 4 $. The nonlinearity of EW is obvious and in this case is a semicircle. We see that the nonlinear EW detects more entangled density matrices.}
	\label{fig:Fig2_PPT_TrWRho_Envelope}
\end{figure}
This is the envelope equation with constraints $0\leq p_{1}+p_{2} \leq 1$. As this is a nonlinear function we call such envelope as \textit{nonlinear entanglement witness}. The detected entangled region (not necessarily $PPT$), and the nonlinear \emph{EW} as the envelope are plotted in Fig.~\ref{fig:Fig2_PPT_TrWRho_Envelope}. As can be seen the nonlinear witness has a wider detection range of entanglement.
\par
As a second example, we provide a state which is clearly $PPT$ across every bi-partition, and which is detected by our nonlinear entanglement witness, (\ref{NonLinearEWsPis}). Consider the following state
\begin{equation}\label{ExplicitPPTState}
\rho=\sum_{i=1}^{8} \frac{1}{16} |\psi_{i}\rangle \langle \psi_{i}|+
\frac{1}{8}(
|\psi_{9}\rangle \langle \psi_{9}|+
|\psi_{11}\rangle \langle \psi_{11}|+
|\psi_{13}\rangle \langle \psi_{13}|+
|\psi_{16}\rangle \langle \psi_{16}|
)
\end{equation}
where we set $p_{1}=\cdots=p_{8}=1/16,\ p_{9}=p_{11}=p_{13}=p_{16}=1/8,\ p_{10}=p_{12}=p_{14}=p_{15}=0$. In matrix form we have
\begin{eqnarray}
\rho=
\left(
\scriptsize{
	\begin{array}{llllllllllllllll}
	1 & 0 & 0 & 0 & 0 & 0 & 0 & 0 & 0 & 0 & 0 & 0 & 0 & 0 & 0 & 0 \\
	0 & 1 & 0 & 0 & 0 & 0 & 0 & 0 & 0 & 0 & 0 & 0 & 0 & 0 & 0 & 0 \\
	0 & 0 & 1 & 0 & 0 & 0 & 0 & 0 & 0 & 0 & 0 & 0 & 0 & 0 & 0 & 0 \\
	0 & 0 & 0 & 1 & 0 & 0 & 0 & 0 & 0 & 0 & 0 & 0 & 0 & 0 & 0 & 0 \\
	0 & 0 & 0 & 0 & 1 & 0 & 0 & 0 & 0 & 0 & 0 & 1 & 0 & 0 & 0 & 0 \\
	0 & 0 & 0 & 0 & 0 & 1 & 0 & 0 & 0 & 0 & 1 & 0 & 0 & 0 & 0 & 0 \\
	0 & 0 & 0 & 0 & 0 & 0 & 1 & 0 & 0 & 1 & 0 & 0 & 0 & 0 & 0 & 0 \\
	0 & 0 & 0 & 0 & 0 & 0 & 0 & 1 & -1 & 0 & 0 & 0 & 0 & 0 & 0 & 0 \\
	0 & 0 & 0 & 0 & 0 & 0 & 0 & -1 & 1 & 0 & 0 & 0 & 0 & 0 & 0 & 0 \\
	0 & 0 & 0 & 0 & 0 & 0 & 1 & 0 & 0 & 1 & 0 & 0 & 0 & 0 & 0 & 0 \\
	0 & 0 & 0 & 0 & 0 & 1 & 0 & 0 & 0 & 0 & 1 & 0 & 0 & 0 & 0 & 0 \\
	0 & 0 & 0 & 0 & 1 & 0 & 0 & 0 & 0 & 0 & 0 & 1 & 0 & 0 & 0 & 0 \\
	0 & 0 & 0 & 0 & 0 & 0 & 0 & 0 & 0 & 0 & 0 & 0 & 1 & 0 & 0 & 0 \\
	0 & 0 & 0 & 0 & 0 & 0 & 0 & 0 & 0 & 0 & 0 & 0 & 0 & 1 & 0 & 0 \\
	0 & 0 & 0 & 0 & 0 & 0 & 0 & 0 & 0 & 0 & 0 & 0 & 0 & 0 & 1 & 0 \\
	0 & 0 & 0 & 0 & 0 & 0 & 0 & 0 & 0 & 0 & 0 & 0 & 0 & 0 & 0 & 1 \\
	\end{array}
}
\right)
\end{eqnarray}
Clearly, this density matrix is $PPT$ across every bi-partition. From (\ref{NonLinearEWsPis}) and choosing $p=1$, we have 
\begin{equation}
Tr(W \rho) = 1 + \sqrt{2} a_{1}
\end{equation}
which for $a_{1}=-1$, yields $1-\sqrt{2}$, then the $PPT$ entangled state can be detected by our nonlinear witness. The reader may note that this state is in the detected region of Fig.~\ref{fig:Fig2_PPT_TrWRho_Envelope} with coordinates $(p_{1}=1/16,\ p_{2}=1/16)$.
\par
Consequently, we achieved the nonlinear \emph{EW}s for some four qubits \emph{MUB} diagonal density matrices and the negativity of (\ref{NonLinearEWsPis}) is the evidence of four qubits entanglement in the system. 
At the end, we present other nonlinear \emph{EW}s families. These have the following form
\begin{eqnarray}\label{64EWs}
W_{ \left\lbrace i_1, i_2, i_3, i_4 \right\rbrace } & = & IIII \pm O_{j}
\nonumber\\&&
+ \ \Big[ \sigma_x\sigma_x\sigma_x\sigma_x + (-1)^{i_{1}} \sigma_x\sigma_x\sigma_y\sigma_y \Big] A_{1}/A_{0}
\nonumber\\&&
+ \ \Big[ \sigma_y\sigma_y\sigma_x\sigma_x + (-1)^{i_{2}} \sigma_y\sigma_y\sigma_y\sigma_y \Big] A_{2}/A_{0}
\nonumber\\&&
+ \ \Big[ \sigma_x\sigma_y\sigma_y\sigma_x + (-1)^{i_{3}} \sigma_x\sigma_y\sigma_x\sigma_y \Big] A_{3}/A_{0}
\nonumber\\&&
+ \ \Big[ \sigma_y\sigma_x\sigma_y\sigma_x + (-1)^{i_{4}} \sigma_y\sigma_x\sigma_x\sigma_y \Big] A_{4}/A_{0} 
\end{eqnarray}
here 
\begin{eqnarray}
\left\lbrace i_1, i_2, i_3, i_4 \right\rbrace \in \left\lbrace  \left\lbrace 0,\ 0,\ 0,\ 0 \right\rbrace ,\ \left\lbrace 0,\ 0,\ 1,\ 1 \right\rbrace,\ \left\lbrace 1,\ 1,\ 0,\ 0 \right\rbrace,\ \left\lbrace 1,\ 1,\ 1,\ 1 \right\rbrace \right\rbrace,
\nonumber
\end{eqnarray}
\begin{eqnarray}
O_{j} \in \left\lbrace \sigma_z\sigma_z II, II\sigma_z\sigma_z, I\sigma_z I\sigma_z, I\sigma_z \sigma_z I,\sigma_z I \sigma_z I, \sigma_z II \sigma_z, \sigma_z \sigma_z \sigma_z \sigma_z  \right\rbrace
\nonumber 
\end{eqnarray}
It is easy to build up $ 2 \times 7 \times 4 = 56 $ nonlinear \emph{EW}s using (\ref{64EWs}), where, $2$ denotes for $\pm$ sign of $Q_{j}$, $7$ denotes for number of elements in $Q_{j}$ set, and $4$ denotes the number of elements in  $\{ i_1, i_2, i_3, i_4  \}$ set. Furthermore, if we consider the notation $ P(m,\ n) $ for permutation of $ m $th and $ n $th Pauli matrices of the eight terms in the brackets of (\ref{64EWs}), then the permutation $ P(1,\ 2),\ P(1,\ 3),\ P(1,\ 4),\ P(2,\ 3)$, and $ P(3,\ 4) $ gives new nonlinear \emph{EW}s. Therefore, we have $ 56 \times 6 = 336 $ nonlinear \emph{EW}s.

\section{Thermal entanglement and its detection}\label{section:6}
Let us consider a canonical ensemble of four qubits identical systems in thermal equilibrium. We would like to find the entanglement detection condition at temperature, $ T $. From equ.~(\ref{NonLinearEWsPis}), one can find the entanglement dependence on the temperature. Rewriting this condition and using equ.~(\ref{pis}) we have (we set $ k_{B} =1 $),
\begin{eqnarray}\label{NonLinearEWsPisTemp}
Tr(W\rho) & = & 
1 + a_{_{0}} \bigg(1-\frac{2}{Z}\sum_{j=9}^{16} e^{-E{_{j}}/T} \bigg)+
\nonumber\\&&
\frac{4 p a_{_{1}}}{Z}
\Bigg[ \Big( e^{\frac{-E{_{11}}}{T}}-e^{\frac{-E{_{12}}}{T}}+e^{\frac{-E{_{13}}}{T}}-e^{\frac{-E{_{14}}}{T}} \Big) ^{2}+\Big(e^{\frac{-E{_{9}}}{T}}-e^{\frac{-E{_{10}}}{T}}-
\nonumber\\&&
e^{\frac{-E{_{15}}}{T}}+e^{\frac{-E{_{16}}}{T}}\Big)^{2}  \Bigg]^{1/2} \textbf{sgn}
\Big(\frac{e^{\frac{-E{_{11}}}{T}}-e^{\frac{-E{_{12}}}{T}}+e^{\frac{-E{_{13}}}{T}}-e^{\frac{-E{_{14}}}{T}}}{Z}\Big)+
\nonumber\\&&
\frac{4 (1-p) a_{_{2}}}{Z}
\Bigg[    \Big( e^{\frac{-E{_{3}}}{T}}-e^{\frac{-E{_{4}}}{T}}+e^{\frac{-E{_{5}}}{T}}-e^{\frac{-E{_{6}}}{T}} \Big) ^{2}+\Big(e^{\frac{-E{_{1}}}{T}}-e^{\frac{-E{_{2}}}{T}}- 
\nonumber\\&&
e^{\frac{-E{_{7}}}{T}}+e^{\frac{-E{_{8}}}{T}}\Big)^{2} \textbf{sgn} \Big(\frac{e^{\frac{-E{_{3}}}{T}}-e^{\frac{-E{_{4}}}{T}}+e^{\frac{-E{_{5}}}{T}}-e^{\frac{-E{_{6}}}{T}}}{Z}\Big)
\end{eqnarray}
in this expression
\begin{equation}\label{Partition_Function}
Z=\sum_{j=1}^{16} e^{-\frac{E{j}}{T}}
\end{equation}
is the partition function of the system. For a given $ E_{j} $s, the negativity of (\ref{NonLinearEWsPisTemp}) for a temperature interval, is the sufficient entanglement condition.
\par
For example, we consider the following  Hamiltonian
\begin{eqnarray}\label{Heisenberg_Like}
H & = & - J \Big( \sigma_{z}\sigma_{z}II + I\sigma_{z}\sigma_{z}I +II\sigma_{z}\sigma_{z} +\sigma_{z}II\sigma_{z}+\sigma_{z}I\sigma_{z}I+I\sigma_{z}I\sigma_{z} \Big) 
\nonumber\\& &
+ h \Big( \sigma_{x}\sigma_{x}\sigma_{x}\sigma_{x} + \sigma_{y}\sigma_{y}\sigma_{y}\sigma_{y} + \sigma_{z}\sigma_{z}\sigma_{z}\sigma_{z} \Big)
\end{eqnarray}
the first part shows the pair coupling between four qubits with coupling constant $ J $, the second part represents some spin interactions among all four qubits with strength $ h $. This Hamiltonian in the Bell-diagonal bases is diagonal with energy eigenvalues
\begin{eqnarray}\label{eigenvalues}
& & E_{1} = 3 h -6 J,\ E_{2}=-h-6J,
\nonumber\\& &
E_{3}=E_{4}=E_{5}=E_{6}=E_{9}=E_{10}=E_{15}=E_{16}= -h,
\nonumber\\& &
E_{7}=E_{11}=E_{13}=3 h +2 J,
\nonumber\\& &
E_{8}=E_{12}=E_{14}=- h +2 J
\end{eqnarray}
and the partition function of the canonical ensemble of four qubits systems at equilibrium temperature, $ T $, is
\begin{equation}\label{PartitionFunctionExample}
Z = 8 e^{ \frac{h}{T} } +3 e^{  \frac{-3h-2J}{T}}  +3 e^{ \frac{h-2J}{T}} + e^{\frac{-3h+6J}{T}} + e^{\frac{h+6J}{T}}
\end{equation} 
Now one can use (\ref{NonLinearEWsPisTemp}) for calculating the entanglement detection condition at temperature $ T $, which is 
\begin{equation}
Tr(W \rho)=\Big[1+ \frac{a_{0} \left(1+e^{\frac{4 h}{T}}\right) \left(-1+e^{\frac{8 J}{T}}\right)+8 a_{1} p \left(1-e^{\frac{4 h}{T}}\right)}{3+3 e^{\frac{4 h}{T}}+8 e^{\frac{2 (2 h+J)}{T}}+\left(1+e^{\frac{4 h}{T}}\right) e^{\frac{8 J}{T}}} \Big] <0
\end{equation}
\begin{figure}
	\centering
	\includegraphics[width=0.7\linewidth]{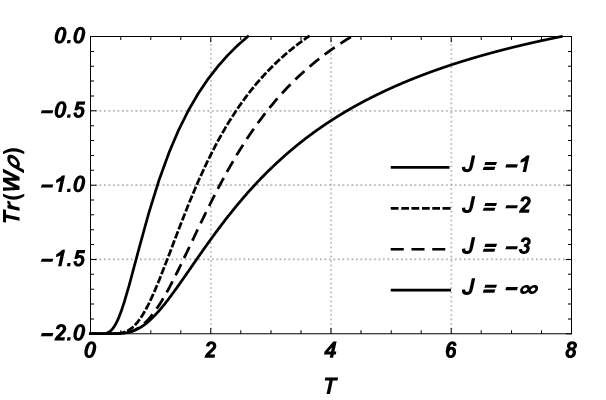}
	\caption[Short Caption]{The entanglement detection in terms of temperature, $ T $ for three values of $ J =-1, -2, -3, -\infty$. Entanglement is detected for $ Tr(W\rho)\leq 0 $. Here we choose $ a_{0}=a_{1}=p=h=1 $. }
	\label{fig:Fig_4}
\end{figure}
Using this inequality we can estimate the threshold temperature, $ T_{th} $, for entanglement in the thermodynamic limit. Fig.~\ref{fig:Fig_4} shows the plot of this condition in terms of temperature for $ a_{0}=a_{1}=p=h=1 $, and three values of $ J =-1, -2, -3 , -\infty$. For $ T<T_{th} $, the entanglement has been detected by our nonlinear witness (\ref{NonLinearEWsPis}). These threshold temperatures are represented in Table.~\ref{tab:1}. It is seen that threshold temperatures increase and reach to their maximum values with the decrease of $ J $. This means for lower $J$, the entanglement can be detected for higher $T$.
\begin{table}
	\caption{Threshold temperatures for Hamiltonian (\ref{Heisenberg_Like}). }
	\label{tab:1}
	\begin{tabular}{lllll}
		\hline\noalign{\smallskip}
		J & -1 & -2 & -3 & -$ \infty $  \\
		\noalign{\smallskip}\hline\noalign{\smallskip}
		$ T_{th} $ & 2.6135 & 3.6232 & 4.3418 & 7.8305 \\
		\noalign{\smallskip}\hline
	\end{tabular}
\end{table}
\par
It is interesting to consider the detection condition for  extreme limits of $ J $ and $ h $,
\begin{equation}\label{LimitTrWRho}
\lim_{h\to\infty,\\J\to-\infty} Tr(W\rho) = \left[  1-\frac{1}{3}(a_{0}+8 a_{1} p) \right] < 0
\end{equation}
or $ (a_{0}+8 a_{1} p) > 3$. This result is independent of temperature and for values such as $ a_{0}=a_{1}=p=1 $, not only the system is entangled in any temperature but also we have a witness detecting this entanglement. 

\par
As the second example of the thermal entanglement detection ability of our witnesses, we consider a four qubits Heisenberg $ XX $ chain in a magnetic field. In \cite{cao2005thermal}, a linear chain Heisenberg $ XX $ model of four qubits in the presence of magnetic field $ B $ is investigated and the pairwise entanglement between alternate qubits is calculated. The four qubits $ XXM $ Heisenberg model is
described by the Hamiltonian
\begin{equation}\label{XXMHamintonian}
H_{XXM} = J \sum_{n=1}^{4} (\sigma_{n}^{+} \sigma_{n+1}^{-} + \sigma_{n}^{-} \sigma_{n+1}^{+} ) + B  \sum_{n=1}^{4} \sigma_{n}^{z}
\end{equation}
where, $ \sigma_{n}^{\pm} $, are the raising and  lowering operators, and $ J $ is the interaction strength. From the energy eigenvalues, $ E_ {I} $, and eigenfunctions, $ | \psi_{i} \rangle $, we can find the density matrix as
\begin{equation}\label{XXMDensityMatrix}
\rho_{XXM} =\frac{1}{Z} \sum_{i=0}^{15} e^{-\beta E_{i}}  | \psi_{i} \rangle \langle \psi_{i} |
\end{equation}
where, $ Z = \sum_{i=0}^{15} e^{-\beta E_{i}} $, is the partition function. Hereafter, we set $ \beta = 1 / T  $. In the original paper, the pairwise entanglement is investigated, but here, we are going to find the global entanglement for this density matrix.  Choosing the following witness, as one of the detecting one,
\begin{eqnarray}\label{Wou3}
W_{2} & = & IIII + \sigma_z\sigma_z II +
\nonumber\\&&
\cos\psi_{_{1}} \left( 
\sigma_x\sigma_x\sigma_x\sigma_x + \sigma_x\sigma_x\sigma_y\sigma_y + \sigma_y\sigma_y\sigma_x\sigma_x + \sigma_y\sigma_y\sigma_y\sigma_y 
\right) 
+
\nonumber\\&&
\sin\psi_{_{1}} \left(  
\sigma_x\sigma_y\sigma_y\sigma_x - \sigma_x\sigma_y\sigma_x\sigma_y +
\sigma_y\sigma_x\sigma_x\sigma_y - \sigma_y\sigma_x\sigma_y\sigma_x 
\right) 
\end{eqnarray}
where we set $ p=1 $, $ O_{j} = \sigma_z\sigma_z II  $, $ i_{1} = i_{2} =0$, $ i_{3}=i_{4} =1 $, in (\ref{64EWs}), we have
\begin{equation}\label{TrRoWCao}
Tr(\rho_{XXM} W_{2}) = K_{1}/K_{2}
\end{equation} 
where
$$
K_{1} = 3 + \cosh \frac{2\sqrt{2}}{T} + 8 \cosh^{2}\frac{1}{T} \cosh \frac{2 B}{T} + 
4 \cosh \frac{4 B}{T} + 4 (3 \cos \psi_{1} + \sin \psi_{1}) \sinh^{2} \frac{\sqrt{2}}{T}
$$
and
$$
K_{2} = 2 \left(4 \cosh^2\left(\frac{1}{T}\right) \cosh \left(\frac{2 B}{T}\right)+\cosh \left(\frac{4 B}{T}\right)+\cosh \left(\frac{2 \sqrt{2}}{T}\right)+2\right)
$$
\begin{figure}
	\centering
	\includegraphics[width=0.7\linewidth]{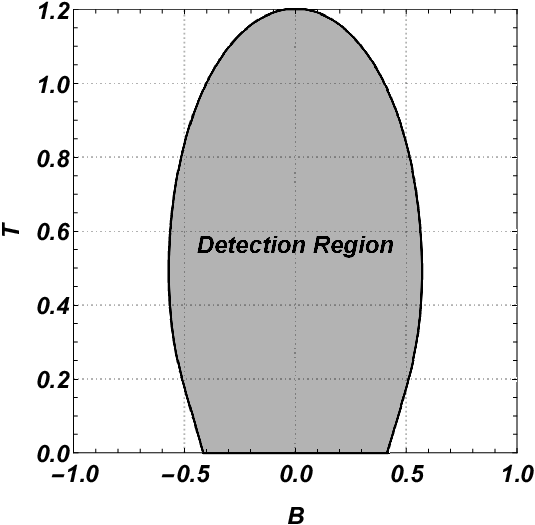}
	\caption[Thermal Example 2]{The global entanglement detection region for (\ref{XXMDensityMatrix}), by our witness, (\ref{Wou3}). Here, we set $ \psi_{1} = 3.46 $.
	}
	\label{Fig_8_R}
\end{figure}
In Fig.~\ref{Fig_8_R}, we give the detection region result in terms of temperature and magnetic field, where we set $ \psi_{1} = 3.46 $. First, we observe that our witness can detect entanglement for wide range of temperature, $ 0 \leq T \leq 1.2 $, with threshold,  $  T_{th} = 1.2 $.  Also, we can see that the thermal state has global entanglement even for $ B = 0 $. 

\par
For a different approach to thermal entanglement and its detection, please see \cite{RefJ:Kay} where the entanglement properties of graph-diagonal states and the linear cluster state are discussed and a relation to the partition function of the classical Ising model is investigated.

\section{Comparison with other results}
For our comparison, we consider two examples, the detection of genuine multipartite entanglement using entanglement witness operators \cite{bourennane2004experimental}, and with the analytical lower bound of concurrence of four qubits mixed quantum sates \cite{zhu2018lower}. 
\par
In \cite{bourennane2004experimental}, the experimental detection of genuine multipartite entanglement using entanglement witness operators is presented. For the four qubits case, the following state is considered 
\begin{equation}\label{bourennane2004experimentalState}
| \psi^{(4)} \rangle = \frac{1}{\sqrt{3}} \left(  | 0011 \rangle +| 1100 \rangle  -\frac{1}{2}  (| 0110 \rangle +| 1001 \rangle +| 0101 \rangle +| 1010 \rangle  )    \right) 
\end{equation}
with the resulting pure density matrix, $ \rho^{(1)} = | \psi^{(4)} \rangle  \langle \psi^{(4)} | $, and the witness, $ W_{ \psi^{(4)}} = \frac{3}{4} \boldmath{I} - | \psi^{(4)} \rangle  \langle \psi^{(4)} | $. This witness detects the entanglement, $ Tr ( W_{ \psi^{(4)}} \rho )  = - \frac{1}{4} $. 
\par 
There are many of our \emph{EW}s detecting this pure state density matrix, for example, in (\ref{64EWs}), with the minus sign in the second term, setting $ O_{j}=\sigma_z\sigma_z II $,  $ p=0 $, and using (\ref{ps1s2}), the witness is

\begin{eqnarray}\label{Wou1}
W_{1} & = & IIII - \sigma_z\sigma_z II +
\nonumber\\&&
\cos\psi_{_{2}} \left( 
\sigma_x\sigma_x\sigma_x\sigma_x  - \sigma_x\sigma_x\sigma_y\sigma_y + \sigma_y\sigma_y\sigma_y\sigma_y  - \sigma_y\sigma_y\sigma_x\sigma_x 
\right) 
+
\nonumber\\&&
\sin\psi_{_{2}} \left(  
\sigma_x\sigma_y\sigma_y\sigma_x + \sigma_x\sigma_y\sigma_x\sigma_y +
\sigma_y\sigma_x\sigma_y\sigma_x + \sigma_y\sigma_x\sigma_x\sigma_y 
\right) 
\end{eqnarray}
with the detection, 
$ Tr(W_{1} \rho^{(1)})  =\frac{2}{3} \left(  1 + 4 ( \sin \psi_{_{2}} + \cos \psi_{_{2}} ) \right)  $ which is negative for 
$ 2 \tan ^{-1}\left(\frac{1}{3} \left(\sqrt{31}+4\right)\right)\leq \psi_{_{2}} \leq 2 \pi +2 \tan ^{-1}\left(\frac{1}{3} \left(4-\sqrt{31}\right)\right) $. 
\par
\begin{figure}
	\centering
	\includegraphics[width=0.7\linewidth]{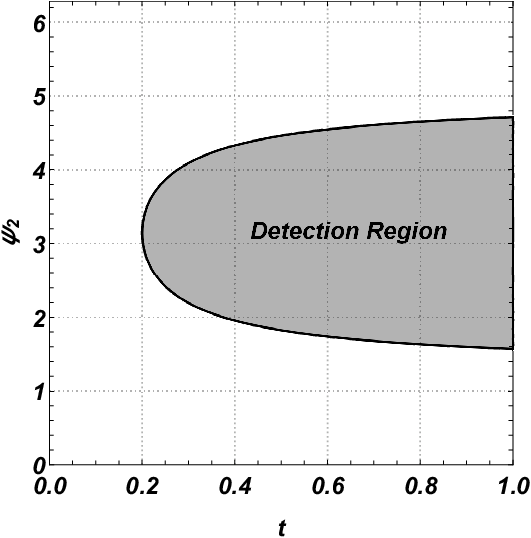}
	\caption[Detection QIP]{Detection Region with our witness, (\ref{Wou1}), for the density matrix given by, \cite{zhu2018lower}. Entanglement is detected for $ \frac{2}{10} \leq t \leq 1 $.}
	\label{fig:fig7r}
\end{figure}
\par
Zho et al., have presented an improved lower bound for multipartite quantum systems in terms of the concurrence , \cite{zhu2018lower}. They also have presented an analytic form for four qubits system. The state and the corresponding mixed state density matrix are as follows 
\begin{equation}\label{ZhuExample}
| \psi_{Zho} \rangle = \frac{1}{2} \left(  | 0000 \rangle +| 0011 \rangle +  | 1100 \rangle  +| 1111 \rangle   \right) 
\end{equation}
and $ \rho_{Zho} = \frac{1-t}{16}  \boldmath{I_{16}}   + t  | \psi_{Zho} \rangle \langle \psi_{Zho} | $. The lower bound is $ t = 1/9 $, which means the entanglement can be detected for $ \frac{1}{9} \leq t \leq 1 $. We examine the entanglement detection ability with our witness, (\ref{Wou1}). The result is
\begin{equation}\label{OurTrWRo42}
Tr (W_{1} \rho_{Zho}) = 1 - t + 4 (1-p) t \cos \psi_{2} 
\end{equation}
which has negative value for, $ t \geq 1 / 5 $. From Fig.~\ref{fig:fig7r}, we can see the entanglement detection region in terms of parameters, $ t $ and $ \psi_{2} $. The entanglement can be detected for $ \frac{2}{10} \leq t \leq 1 $ while $ \psi_{2} $ varies from $ \pi $ to $ 3 \pi /2 $. Except for, entanglement detection in \cite{zhu2018lower}, our results are better than the results of \cite{li2015genuine} and \cite{zhu2014lower} for this density matrix, which are discussed in \cite{zhu2018lower}.  The concurrence in \cite{zhu2018lower}, detects mixed state entanglement for $ \frac{1}{9}  < t \leq \frac{2}{10}$, where, our witnesses fail to detect. 
\section{Conclusion}
We have constructed linear and nonlinear entanglement witnesses with a wider detection region for four qubits systems in mutually unbiased bases for a given diagonal Hamiltonian in those bases. These witnesses can detect the entangled positive partial transpose density matrices. We established the envelope of a family of linear witnesses as a nonlinear witness. We applied them to detect the thermal entanglement in a canonical ensemble with an infinite number of four qubits in thermal equilibrium at temperature, $ T $, and we demonstrated that these witnesses can detect the thermal entanglement for some Hamiltonians even at any temperature. Our results highlight the potential of this method and may be applied to the entanglement investigation of systems for multi-qubits such as the Heisenberg model.

\appendix

\section{four qubits $MUB$ density matrix in terms of Pauli matrices}
It is useful to present the four qubits $MUB$ density matrix in terms of the Pauli matrices 
\begin{eqnarray}
\rho& = &\frac{1}{16} \Big(IIII + r_{_{1}} I\sigma_z\sigma_zI + r_{_{2}} I\sigma_zI\sigma_z + r_{_{3}} II\sigma_z\sigma_z + 
r_{_{4}}\sigma_z I I \sigma_z+ 
\nonumber\\&&
r_{_{5}}\sigma_zI\sigma_zI + 
r_{_{6}}\sigma_z\sigma_zII+
r_{_{7}}\sigma_z\sigma_z\sigma_z\sigma_z +
r_{_{8}}\sigma_x\sigma_x\sigma_x\sigma_x+
\nonumber\\&&
r_{_{9}}\sigma_x\sigma_y\sigma_y\sigma_x +
r_{_{_{10}}}\sigma_x\sigma_y\sigma_x\sigma_y +
r_{_{_{11}}}\sigma_x\sigma_x\sigma_y\sigma_y + r_{_{_{12}}}\sigma_y\sigma_x\sigma_y\sigma_x +
\nonumber\\&&
r_{_{_{13}}}\sigma_y\sigma_x\sigma_x\sigma_y +
r_{_{_{14}}}\sigma_y\sigma_y\sigma_x\sigma_x + r_{_{_{15}}}\sigma_y\sigma_y\sigma_y\sigma_y \Big)
\end{eqnarray}
where the coefficients are
\begin{eqnarray}
r_{_{1}}\;\,& = & p_{_{1}}+p_{_{2}}+p_{_{3}}+p_{_{4}}-p_{_{5}}-p_{_{6}}-p_{_{7}}-p_{_{8}}- p_{_{9}}-p_{_{10}}
\nonumber\\&&
-p_{_{11}}-p_{_{12}}+p_{_{13}}+p_{_{14}}+p_{_{15}}+p_{_{16}}
\nonumber\\
r_{_{2}}\;\,& = & p_{_{1}}+p_{_{2}}-p_{_{3}}-p_{_{4}}+p_{_{5}}+p_{_{6}}-p_{7}-p_{_{8}}-p_{_{9}}-p_{_{10}}
\nonumber\\&&
+p_{_{11}}+p_{_{12}}-p_{_{13}}-p_{_{14}}+p_{_{15}}+p_{_{16}}
\nonumber\\
r_{_{3}}\;\,& = & p_{_{1}}+p_{_{2}}-p_{_{3}}-p_{_{4}}-p_{_{5}}-p_{_{6}}+p_{_{7}}+p_{_{8}}+p_{_{9}}+p_{_{10}}
\nonumber\\&&
-p_{_{11}}-p_{_{12}}-p_{_{13}}-p_{_{14}}+p_{_{15}}+p_{_{16}}
\nonumber\\
r_{_{4}}\;\,& = & p_{_{1}}+p_{_{2}}-p_{_{3}}-p_{_{4}}+p_{_{5}}+p_{_{6}}-p_{_{7}}-p_{_{8}}+p_{_{9}}+p_{_{10}}
\nonumber\\&&
-p_{_{11}}-p_{_{12}}+p_{_{13}}+p_{_{14}}-p_{_{15}}-p_{_{16}}
\nonumber\\
r_{_{5}}\;\,& = & p_{_{1}}+p_{_{2}}+p_{_{3}}+p_{_{4}}-p_{_{5}}-p_{_{6}}-p_{_{7}}-p_{_{8}}+p_{_{9}}+p_{_{10}}
\nonumber\\&&
+p_{_{11}}+p_{_{12}}-p_{_{13}}-p_{_{14}}-p_{_{15}}-p_{_{16}}
\nonumber\\
r_{_{6}}\;\,& = & p_{_{1}}+p_{_{2}}+p_{_{3}}+p_{_{4}}+p_{_{5}}+p_{_{6}}+p_{_{7}}+p_{_{8}}-p_{_{9}}-p_{_{10}}
\nonumber\\&&
-p_{_{11}}-p_{_{12}}-p_{_{13}}-p_{_{14}}-p_{_{15}}-p_{_{16}}
\nonumber\\
r_{_{7}}\;\,& = & p_{_{1}}+p_{_{2}}-p_{_{3}}-p_{_{4}}-p_{_{5}}-p_{_{6}}+p_{_{7}}+p_{_{8}}-p_{_{9}}-p_{_{10}}
\nonumber\\&&
+p_{_{11}}+p_{_{12}}+p_{_{13}}+p_{_{14}}-p_{_{15}}-p_{_{16}}
\nonumber\\
r_{_{8}}\;\,& = & p_{_{1}}-p_{_{2}}+p_{_{3}}-p_{_{4}}+p_{_{5}}-p_{_{6}}+p_{_{7}}-p_{_{8}}+p_{_{9}}-p_{_{10}}
\nonumber\\&&
+p_{_{11}}-p_{_{12}}+p_{_{13}}-p_{_{14}}+p_{_{15}}-p_{_{16}}
\nonumber\\
r_{_{9}}\;\,& = & -p_{_{1}}+p_{_{2}}-p_{_{3}}+p_{_{4}}+p_{_{5}}-p_{_{6}}+p_{_{7}}-p_{_{8}}+p_{_{9}}-p_{_{10}}
\nonumber\\&&
+p_{_{11}}-p_{_{12}}-p_{_{13}}+p_{_{14}}-p_{_{15}}+p_{_{16}}
\nonumber\\
r_{_{_{10}}}& = & -p_{_{1}}+p_{_{2}}+p_{_{3}}-p_{_{4}}-p_{_{5}}+p_{_{6}}+p_{_{7}}-p_{_{8}}+p_{_{9}}-p_{_{10}}
\nonumber\\&&
-p_{_{11}}+p_{_{12}}+p_{_{13}}-p_{_{14}}-p_{_{15}}+p_{_{16}}
\nonumber\\
r_{_{_{11}}}& = & -p_{_{1}}+p_{_{2}}+p_{_{3}}-p_{_{4}}+p_{_{5}}-p_{_{6}}-p_{_{7}}+p_{_{8}}-p_{_{9}}+p_{_{10}}
\nonumber\\&&
+p_{_{11}}-p_{_{12}}+p_{_{13}}-p_{_{14}}-p_{_{15}}+p_{_{16}}
\nonumber\\
r_{_{_{12}}}& = & -p_{_{1}}+p_{_{2}}-p_{_{3}}+p_{_{4}}+p_{_{5}}-p_{_{6}}+p_{_{7}}-p_{_{8}}-p_{_{9}}+p_{_{10}}
\nonumber\\&&
-p_{_{11}}+p_{_{12}}+p_{_{13}}-p_{_{14}}+p_{_{15}}-p_{_{16}}
\nonumber\\
r_{_{_{13}}}& = & -p_{_{1}}+p_{_{2}}+p_{_{3}}-p_{_{4}}-p_{_{5}}+p_{_{6}}+p_{_{7}}-p_{_{8}}-p_{_{9}}+p_{_{10}}
\nonumber\\&&
+p_{_{11}}-p_{_{12}}-p_{_{13}}+p_{_{14}}+p_{_{15}}-p_{_{16}}
\nonumber\\
r_{_{_{14}}}& = & -p_{_{1}}+p_{_{2}}-p_{_{3}}+p_{_{4}}-p_{_{5}}+p_{_{6}}-p_{_{7}}+p_{_{8}}+p_{_{9}}-p_{_{10}}
\nonumber\\&&
+p_{_{11}}-p_{_{12}}+p_{_{13}}-p_{_{14}}+p_{_{15}}-p_{_{16}}
\nonumber\\
r_{_{_{15}}}& = & p_{_{_{1}}}-p_{_{_{2}}}-p_{_{_{3}}}+p_{_{_{4}}}-p_{_{_{5}}}+p_{_{_{6}}}+p_{_{_{7}}}-p_{_{_{8}}}-p_{_{_{9}}}+p_{_{_{10}}}
\nonumber\\&&
+p_{_{_{11}}}-p_{_{_{12}}}+p_{_{_{13}}}-p_{_{_{14}}}-p_{_{_{15}}}+p_{_{_{16}}}
\nonumber
\end{eqnarray}

\section{Two cases for $FR$}
In order to better visualization of the feasible region, here we present the two special cases as follows.
\subsection{The first case }
Let us consider the $(p_{_{1}},p_{_{2}})$ and $(p_{_{3}},p_{_{4}})$ planes. From (\ref{p1p2Inequality_1}), and similar calculations for $ (p_{3},\ p_{4}) $ pair, we have the following inequalities
\begin{equation}
\begin{array}{l}
8p_{_{1}}-6p_{_{2}}\leq 1\\
8p_{_{3}}-6p_{_{4}}\leq 1
\end{array}
\end{equation}
We can find the $PPT$ region boundaries in the $(p_{_{1}},p_{_{2}},p_{_{3}},p_{_{4}})$ space by requiring that
\begin{equation}
\left\{\begin{array}{l}
8p_{_{1}}-6p_{_{2}}=1\\
8p_{_{3}}-6p_{_{4}}=1
\end{array}
\right.\Rightarrow 
\left\{\begin{array}{l}
p_{_{1}}=3 p_{_{2}}/4 +1/8\\
p_{_{3}}=3 p_{_{4}}/4 +1/8
\end{array}\right.
\end{equation}
\par
Similarly the $PPT$ conditions for following cases 
\[ \begin{array}{lclclcl}
(p_{_{1}},p_{_{2}},\;p_{_{5}},\;p_{_{6}})&,& 
(p_{_{1}},p_{_{2}},\;p_{_{7}},\;p_{_{8}})&,& 
(p_{_{1}},p_{_{2}},\;p_{_{9}},p_{_{10}})&,&  (p_{_{1}},p_{_{2}},p_{_{11}},p_{_{12}}),\\ 
(p_{_{1}},p_{_{2}},p_{_{13}},p_{_{14}})&,&
(p_{_{1}},p_{_{2}},p_{_{15}},p_{_{16}})&,& 
(p_{_{3}},p_{_{4}},\;p_{_{5}},\,\,p_{_{6}})&,& 
(p_{_{3}},p_{_{4}},\;p_{_{7}},\,\,p_{_{8}}),\\
(p_{_{3}},p_{_{4}},\,p_{_{9}},\,p_{_{10}})&,& 
(p_{_{3}},p_{_{4}},p_{_{11}},p_{_{12}})&,& 
(p_{_{3}},p_{_{4}},p_{_{13}},p_{_{14}})&,&
(p_{_{3}},p_{_{4}},p_{_{15}},p_{_{16}})
\end{array}\]
are satisfied unless for the following cases
\begin{equation}
\left\{
\begin{array}{lcr}
-p_{_{1}}+p_{_{2}}+p_{_{5}}+p_{_{6}}&\geq& 0\\
-p_{_{1}}+p_{_{2}}+p_{_{7}}+p_{_{8}}&\geq& 0\\
-p_{_{1}}+p_{_{2}}+p_{_{9}}+p_{_{10}}&\geq& 0\\
-p_{_{1}}+p_{_{2}}+p_{_{11}}+p_{_{12}}&\geq& 0\\
-p_{_{1}}+p_{_{2}}+p_{_{13}}+p_{_{14}}&\geq& 0\\
-p_{_{1}}+p_{_{2}}+p_{_{15}}+p_{_{16}}&\geq& 0\\
-p_{_{3}}+p_{_{4}}+p_{_{5}}+p_{_{6}}&\geq& 0\\
-p_{_{3}}+p_{_{4}}+p_{_{7}}+p_{_{8}}&\geq& 0\\ -p_{_{3}}+p_{_{4}}+p_{_{9}}+p_{_{10}}&\geq& 0\\
-p_{_{3}}+p_{_{4}}+p_{_{11}}+p_{_{12}}&\geq& 0\\
-p_{_{3}}+p_{_{4}}+p_{_{13}}+p_{_{14}}&\geq& 0\\
-p_{_{3}}+p_{_{4}}+p_{_{15}}+p_{_{16}}&\geq& 0
\end{array}\right.\Rightarrow
\left\{
\begin{array}{lcr}
p_{_{5}}+p_{_{6}}+p_{_{2}}/4&\geq& 1/8\\
p_{_{7}}+p_{_{8}}+p_{_{2}}/4&\geq& 1/8\\
p_{_{9}}+p_{_{10}}+p_{_{2}}/4&\geq& 1/8\\
p_{_{11}}+p_{_{12}}+p_{_{2}}/4&\geq& 1/8\\
p_{_{13}}+p_{_{14}}+p_{_{2}}/4&\geq& 1/8\\
p_{_{15}}+p_{_{16}}+p_{_{2}}/4&\geq& 1/8\\
p_{_{5}}+p_{_{6}}+p_{_{4}}/4&\geq& 1/8\\
p_{_{7}}+p_{_{8}}+p_{_{4}}/4&\geq& 1/8\\
p_{_{9}}+p_{_{10}}+p_{_{4}}/4&\geq& 1/8\\
p_{_{11}}+p_{_{12}}+p_{_{4}}/4&\geq& 1/8\\
p_{_{13}}+p_{_{14}}+p_{_{4}}/4&\geq& 1/8\\
p_{_{15}}+p_{_{16}}+p_{_{4}}/4&\geq& 1/8
\end{array}\right.
\end{equation}
Adding the above inequalities yield to
\begin{eqnarray}
2(p_{_{5}}+p_{_{6}}+p_{_{7}}+p_{_{8}}+p_{_{9}}+p_{_{10}}+
p_{_{11}}+p_{_{12}}+p_{_{13}}+p_{_{14}}+p_{_{15}}+p_{_{16}})
\nonumber\\
+\frac{3}{2}(p_{_{2}}+p_{_{4}})\geq\frac{3}{2}
\end{eqnarray}
and using the normalization condition for $p_i$'s yields
\begin{equation}
2[1-(p_{_{1}}+p_{_{2}}+p_{_{3}}+p_{_{4}})]+\frac{3}{2}(p_{_{2}}+p_{_{4}})\geq\frac{3}{2}\Rightarrow-\frac{1}{2}(p_{_{2}}+p_{_{4}})\geq 0
\end{equation}
so we must have $p_{_{2}}=p_{_{4}}=0 \Rightarrow p_{_{1}}=p_{_{3}}=\frac{1}{8}$. So the $ PPT $ conditions take the following simpler form
\begin{equation}
\left\{\begin{array}{lcr}
p_{_{5}}\;+p_{_{6}}&\geq& 1/8\\
p_{_{7}}\;+p_{_{8}}&\geq& 1/8\\
p_{_{9}}\;+p_{_{10}}&\geq& 1/8\\
p_{_{11}}+p_{_{12}}&\geq& 1/8\\
p_{_{13}}+p_{_{14}}&\geq& 1/8\\
p_{_{15}}+p_{_{16}}&\geq& 1/8
\end{array}\right.
\end{equation}
Also this is a special case, but now the $112$, $PPT$ inequalities reduced to only six simple inequalities which one can concern it easily.
\subsection{The second case}
Similar to the previous case and using the results, for this case we consider $p_{_{2}}=0 $ and $8p_{_{3}}-6p_{_{4}}=1$, then
\begin{equation}\label{CaseII_inequ_p1p4}
p_{_{1}}-p_{_{3}}+p_{_{4}}\geq0\Rightarrow p_{_{1}}+\frac{p_{_{4}}}{4}\geq\frac{1}{8}
\end{equation}
So the $PPT$ conditions for $(p_{_{1}},p_{_{2}},p_{_{3}},p_{_{4}})$ are satisfied.\\
Adding the following PPT conditions
\begin{equation}
\left\{\begin{array}{lcr}
-p_{_{3}}+p_{_{4}}+p_{_{5}}\;+p_{_{6}}&\geq& 0\\
-p_{_{3}}+p_{_{4}}+p_{_{7}}\;+p_{_{8}}&\geq& 0\\
-p_{_{3}}+p_{_{4}}+p_{_{9}}\;+p_{_{10}}&\geq& 0\\
-p_{_{3}}+p_{_{4}}+p_{_{11}}+p_{_{12}}&\geq& 0\\ -p_{_{3}}+p_{_{4}}+p_{_{13}}+p_{_{14}}&\geq& 0\\ -p_{_{3}}+p_{_{4}}+p_{_{15}}+p_{_{16}}&\geq& 0
\end{array}\right.
\Rightarrow \left\{\begin{array}{lcr}
p_{_{5}}\;+p_{_{6}}\;+p_{_{4}}/4&\geq& 1/8\\
p_{_{7}}\;+p_{_{8}}\;+p_{_{4}}/4&\geq& 1/8\\ p_{_{9}}\;+p_{_{10}}+p_{_{4}}/4&\geq& 1/8\\
p_{_{11}}+p_{_{12}}+p_{_{4}}/4&\geq& 1/8\\ p_{_{13}}+p_{_{14}}+p_{_{4}}/4&\geq& 1/8\\
p_{_{15}}+p_{_{16}}+p_{_{4}}/4&\geq& 1/8
\end{array}\right.
\end{equation}
gives
\begin{equation}\label{ineq_p5p16p5}
p_{_{5}}+...+p_{_{16}}+\frac{3}{2}p_{_{4}}\geq\frac{3}{4}
\end{equation} 
Now if we set $p_{_{1}}=\frac{1}{8}-\frac{p_{_{4}}}{4}+\epsilon$, where $\epsilon \geq0$, the normalization condition, $ \sum_{i=1}^{16}p_i =1 $, and $p_2=0,\ 8p_3 -6p_4 =1 $ yield to
\begin{eqnarray}
\epsilon+\frac{3}{2}p_{_{4}}+p_{_{5}}+p_{_{6}}+...+p_{_{16}}=\frac{3}{4}
\end{eqnarray}
By applying (\ref{ineq_p5p16p5}), we have $-\epsilon\geq0\Rightarrow \epsilon=0$, and   $p_{_{1}}=\frac{1}{8}-\frac{p_{_{4}}}{4}$. Finally 
\begin{equation}
\left\{\begin{array}{lcr}
p_{_{5}}\;+p_{_{6}}\;+p_{_{4}}/4 &=& 1/8\\
p_{_{7}}\;+p_{_{8}}\;+p_{_{4}}/4 &=& 1/8\\ 
p_{_{9}}\;+p_{_{10}}+p_{_{4}}/4 &=& 1/8\\
p_{_{11}}+p_{_{12}}+p_{_{4}}/4 &=& 1/8\\ 
p_{_{13}}+p_{_{14}}+p_{_{4}}/4 &=& 1/8\\
p_{_{15}}+p_{_{16}}+p_{_{4}}/4 &=& 1/8\\
\end{array}\right.\Rightarrow 
\left\{\begin{array}{lllll}
p_{_{1}}&=&p_{_{5}}+p_{_{6}}&=& p_{_{7}}\;+p_{_{8}}\\
&=& p_{_{9}}+p_{_{10}}&=& p_{_{11}}+p_{_{12}}\\ 
&=& p_{_{13}}+p_{_{14}}&=& p_{_{15}}+p_{_{16}}
\end{array}\right.
\end{equation} 

\end{document}